\documentclass{aa}  

\usepackage{graphicx, hyperref, longtable, caption, lastpage}
\usepackage[normalem]{ulem}
\hypersetup{colorlinks = true,
            citecolor = {blue},
            }

\usepackage{txfonts}

\newcommand{\hi}{H\,{\textsc{i}}}

\newcommand{\oiii}{O\,{\textsc{iii}}}
\newcommand{\oii}{O\,{\textsc{ii}}}
\newcommand{\cii}{C\,{\textsc{ii}}}
\newcommand{\neiii}{Ne\,{\textsc{iii}}}
\newcommand{\niv}{N\,{\textsc{iv}}}
\newcommand{\civ}{C\,{\textsc{iv}}}
\newcommand{\heii}{He\,{\textsc{ii}}}
\newcommand{\niii}{N\,{\textsc{iii}}}
\newcommand{\ciii}{C\,{\textsc{iii}}}
\newcommand{\lya}{Ly$\alpha$}
\newcommand{\jwst}{{\em JWST}}
\newcommand{\alma}{{\em ALMA}}

\begin{document} 
   \title{Characterising Ly$\alpha$ damping wings at the onset of reionisation}
   \subtitle{Evidence for highly efficient star formation driven by dense, neutral gas in UV-bright galaxies at $z>9$}
    \titlerunning{Ly$\alpha$ damping wings in galaxies at the onset of reionisation}
    \authorrunning{Pollock et al.}

   \author{Clara L. Pollock\thanks{E-mail: clara.pollock@nbi.ku.dk}\inst{1}\fnmsep\inst{2} 
          \and
          Kasper E. Heintz\inst{3}\fnmsep\inst{1}\fnmsep\inst{2}
          \and
          Joris Witstok\inst{1}\fnmsep\inst{2} 
          \and
          Rashmi Gottumukkala\inst{1}\fnmsep\inst{2} 
          \and
          Gabriel Brammer\inst{1}\fnmsep\inst{2} 
          \and
          Sownak Bose\inst{4}
          \and
          Alex J. Cameron\inst{1}\fnmsep\inst{2} 
          \and
          Pratika Dayal\inst{5, 6, 7}
          \and
          Pieter van Dokkum\inst{8}
          \and
          Johan Fynbo\inst{1}\fnmsep\inst{2} 
          \and
          Viola Gelli\inst{1}\fnmsep\inst{2} 
          \and 
          Matthew J. Hayes\inst{9}
          \and
          Akio K. Inoue\inst{10}\fnmsep\inst{11}
          \and
          Claudia del P. Lagos\inst{12}\fnmsep\inst{1}
          \and
          Peter Laursen
          \inst{1}\fnmsep\inst{2} 
          \and 
          Romain A. Meyer\inst{13}
          \and
          Rohan Naidu\inst{14}
          \and
          Pascal Oesch\inst{13, 1, 2}
          \and
          Lucie E. Rowland\inst{15}
          \and
          Nial R. Tanvir\inst{16}
          \and
          Sandro Tacchella\inst{17}\fnmsep\inst{18}
          \and
          Chamilla Terp
          \inst{1}\fnmsep\inst{2} 
          \and 
          Francesco Valentino
          \inst{3}\fnmsep\inst{1}
          \and
          Fabian Walter\inst{19}
          \and
          John Weaver\inst{14}
          \and
          Callum Witten\inst{13}
          }
   \institute{Cosmic Dawn Center (DAWN), Denmark
         \and Niels Bohr Institute, University of Copenhagen, Jagtvej 128, 2200 Copenhagen N, Denmark
    \and DTU Space, Technical University of Denmark, Elektrovej 327, DK2800 Kgs. Lyngby, Denmark
    \and Institute for Computational Cosmology, Department of Physics, Durham University, South Road, Durham DH1 3LE, UK
    \and Canadian Institute for Theoretical Astrophysics, 60 St George St, University of Toronto, Toronto, ON M5S 3H8, Canada 
    \and David A. Dunlap Department of Astronomy and Astrophysics, University of Toronto, 50 St George St, Toronto ON M5S 3H4, Canada 
    \and Department of Physics, 60 St George St, University of Toronto, Toronto, ON M5S 3H8, Canada
    \and Department of Astronomy, Yale University, New Haven, CT 06511, USA
    \and Stockholm University, Department of Astronomy and Oskar Klein Centre for Cosmoparticle Physics, AlbaNova University Centre, SE-10691, Stockholm, Sweden
    \and Department of Physics, School of Advanced Science and Engineering, Faculty of Science and Engineering, Waseda University, 3-4-1 Okubo, Shinjuku, Tokyo 169-8555, Japan
    \and Waseda Research Institute for Science and Engineering, Faculty of Science and Engineering, Waseda University, 3-4-1 Okubo, Shinjuku, Tokyo 169-8555, Japan
    \and 
    International Centre for Radio Astronomy Research (ICRAR), M468, University of Western Australia, 35 Stirling Hwy, Crawley, WA 6009, Australia
    \and
    Department of Astronomy, University of Geneva, Chemin Pegasi 51, 1290 Versoix, Switzerland
    \and MIT Kavli Institute for Astrophysics and Space Research, 70 Vassar
    Street, Cambridge, MA 02139, USA
    \and Leiden Observatory, Leiden University, PO Box 9513, NL-2300 RA Leiden,
    The Netherlands
    \and School of Physics and Astronomy, University of Leicester, University Rd, Leicester LE1 7RH, United Kingdom
    \and 
    Kavli Institute for Cosmology, University of Cambridge, Madingley Road, Cambridge CB3 0HA, UK
    \and Cavendish Laboratory, University of Cambridge, 19 JJ Thomson Avenue, Cambridge CB3 0HE, UK
    \and Max-Planck-Institut für Astronomie, Königstuhl 17, D-69117 Heidelberg, Germany
\\}

   \date{Received 2026; accepted XX; published XX}

  \abstract 
  {One of the major conundrums in contemporary extragalactic astrophysics is the apparent overabundance of a remarkable population of UV-bright galaxies at redshifts $z\gtrsim 9$. 
  We analyse 
  galaxies spectroscopically observed by \jwst/NIRSpec Prism and confirmed to lie at $z>9$, with sufficient signal-to-noise to carefully model their rest-frame UV to optical continua and line emission. In particular, we model the damped Lyman-$\alpha$ (Ly$\alpha$) absorption (DLA) features of each galaxy to place observational constraints on the gas assembly of neutral atomic hydrogen (\hi) onto the galaxy halos at the onset of cosmic reionisation. 
  Based on the derived \hi\ column densities and star-formation rate (SFR) surface densities, we show that all galaxies are highly efficient at forming stars on rapid $\sim 10-100\,$Myr depletion timescales, greatly in excess compared to the canonical local universe Kennicutt-Schmidt relation and predictions from state-of-the-art galaxy formation simulations. The dense \hi\ gas appears to also drive the offset from the fundamental-metallicity relation of these galaxies though its dust-to-gas ratio is seemingly consistent with values derived for local galaxies except for the lowest metallicity sight-lines. 
  Our results provide the first robust observational constraints on the impact of pristine \hi\ gas on early galaxy assembly, and imply that a combination of highly efficient star formation and low dust obscuration can likely explain the UV-brightness of galaxies at cosmic dawn. 
  }


   \maketitle

\clearpage

\section{Introduction}
The formation of the first stars and galaxies at cosmic dawn is widely thought to be initiated by the infall of pristine neutral gas onto dark matter halos \citep[][]{White78,Keres05,Dekel09,Schaye10,Dayal18}.
Early observations of \jwst\ revealed that the first galaxies at $z\gtrsim10$ were far more abundant and luminous than previously expected \citep[e.g.][]{Castellano22, Naidu22, Atek23a, Bouwens23, Donnan23, Finkelstein23, Harikane23, Adams24, McLeod24, Robertson24, Carniani24a}. 
Further, these most distant galaxies appear to have lower chemical abundances than expected from their mass and star-formation rates \citep[SFRs; e.g.][]{Heintz23_FMR,Curti24,Nakajima23,Pollock25}. 
A key missing piece of these contemporary puzzles is the abundance and distribution of neutral atomic hydrogen (\hi), how it drives early star formation, impacts local galaxy formation prescriptions, and the escape of ionising photons from the galaxies to the large-scale intergalactic medium (IGM).

A direct measure of \hi\ in the high-redshift universe has typically been determined from Lyman-$\alpha$ absorption in bright background objects like quasars \citep{White03, Wolfe05, Fan06} or gamma-ray bursts \citep{Totani06,Prochaska07,Fynbo09,Tanvir19}. These trace the neutral hydrogen on various scales from the interstellar and circumgalactic media (ISM and CGM, respectively) to the IGM, but only for a narrow line of sight, and with few measurements existing beyond redshift $z\sim 6.5$. However, early studies of the UV-turnover in JWST spectra revealed several strong damped \lya\ absorption (DLA) systems with column densities $N_{\rm HI} > 10^{22}\, \rm cm^{-2}$ \citep{Heintz24_DLA, Heintz25, DEugenio24, chen24, Umeda24}, implying absorption far greater than would be expected from gas in the IGM alone. These damped wings are evidence of significant neutral gas which is associated with the galaxy itself. 

Including DLAs in modelling early galaxy spectra is important for obtaining accurate spectroscopic redshifts \citep[][Heintz et al. in prep]{Witstok25_ALMA,Asada25}. They can also have a profound impact on reionisation studies; the gas is fully self-shielding and thereby prevents escape of ionising photons. Understanding the distribution and covering fraction is thus vital for early reionisation models and for accurately measuring the neutral hydrogen fraction of the IGM, $x_\mathrm{HI}$ \citep{Huberty25, Mason26, Umeda24, Umeda26}.

Originally it was hypothesised that the galaxy DLAs observed could be probing the infalling pristine gas itself; which would be diluting the metal-enriched gas of the ISM, possibly explaining the offset observed from the fundamental metallicity relation (FMR) at high-z \citep[][]{Heintz23_FMR, Nakajima23,Curti24,Pollock25}. However, recent cosmological zoom-in simulation studies \citep{Gelli25} have suggested that for most lines of sight, the \hi\ gas column density is dominated by the ISM and central starburst regions, with minimal chance of intersecting a pristine filament or accretion channel. Further, \cite{Rowland25_DLA} find tentative evidence that the neutral gas content inferred from [\cii] luminosities is correlated to the \hi\ gas mass derived from the damping wings, indicating that DLAs are likely associated with gas in the star-forming regions of the ISM. The gas traced by galaxy DLAs at high-redshifts therefore likely provide new and unique insights to many of the processes governing galaxy growth and the `baryon cycle' including star-formation, chemical enrichment, and dust production from the available inflow of pristine, neutral gas. 

The luminous galaxies at $z>10$ are particularly puzzling in this framework.
Several explanations have been proposed to alleviate their tension with galaxy formation models, including bursty star formation \citep{Mason23, Shen23, Gelli24}, a top-heavy initial mass function (IMF) \citep{Harikane24, Menon24, Rasmussen24, Hutter25, Mauerhofer25}, or enhanced star-formation efficiency from feedback-free systems \citep{Dekel23, BoylanKolchin25}. Given the high star-formation rates and compact sizes of the earliest galaxies \citep{RobertsBorsani25,Naidu25,Tang25}, one would expect short free-fall and depletion times, leading to efficient gas-to-star conversion. Moreover, a short free-fall time implies limited dust production \citep{Asano13}, which would further contribute to the exceptional brightness of these primordial systems \citep{Ferrara23}.

The visibility of prominent Ly$\alpha$ emission even in some of the most distant galaxies \citep[e.g. at $z\approx 11$ and $z\approx 13$;][]{Bunker23_gnz11,Witstok25_LA}, at a point in cosmic time where the IGM is expected to be fully neutral, is similarly surprising. Supporting these observations is the detection of a mild Ly$\alpha$ damping in MoM-z14 \citep{Naidu25}, implying a non-negligible fraction of ionised hydrogen in the IGM already at $z\gtrsim 14$. These features can potentially be explained by significant outflows clearing the gas locally and thus allowing Ly$\alpha$ and Lyman-Continuum (LyC) photons to escape \citep{Ferrara24a, Witten24}. Or, more intriguingly, it may constitute the first hints of an increased optical depth to reionisation \citep{Sailer25}, suggestion an earlier onset of reionisation.

In this work, we compile all publicly available JWST spectra of galaxies at $z>9$ to investigate the physical origin and properties of the discovered UV-bright galaxy population. The main goals are to constrain the earliest phases of \hi\ gas mass assembly, how this drives the elevated star-formation efficiencies, and impacts the onset of reionisation. This is now possible via the observed strong DLA features and implied gas column densities in this most distant galaxy population.

Our paper is structured as follows: in Section \ref{sec:observations} we briefly describe the data reduction, detail our sample selection, and introduce the observations. In Section \ref{sec:analysis}, we outline the methodology used to model damped \lya\ absorption, emission lines, as well as calculations for star-formation rate and other gas properties. In Section \ref{sec:results1} we present the distribution of neutral gas, dust, and metals in our sample and compare to simulations. 
Finally, in \ref{sec:conclusion} we summarise and conclude on our work.
Throughout this paper, we assume standard cosmology derived from the Astropy cosmology package, with cosmological parameters from \cite{Planck18}, and solar abundance $\mathrm{12+log(O/H)_\odot} = 8.69$ \citep{Asplund09}.

\section{Observations} \label{sec:observations}

For this work, we compile all galaxies with a robust spectroscopic redshift (grade 3) at $z>9$, and sufficient S/N to model the rest-frame UV from the DAWN JWST Archive \citep[DJA;][]{Brammer_DJA} \texttt{version 4} \citep{Valentino25, Pollock25}, (previously \cite{Heintz25} for \texttt{v2}, \cite{DeGraaff25} for \texttt{v3}) with full JWST/NIRSpec Prism spectroscopy \citep[$\lambda = 0.6 - 5.5 \mu \mathrm{m}$, $R \approx 100$, $R(1.3\mu \rm{m}) \sim 30$;][]{Jakobsen22}. The above references detail the updates from the default \jwst\ reduction pipeline for the Mikulski Archive for Space Telescope (MAST) data products. The most important update for the v4 spectra used here is the bar shadow correction, improving the absolute and colour-dependent flux calibration, and the extended wavelength range in the red end from $5.3$ (default) to $5.5\mu \mathrm{m}$. We further only consider sources with a median signal-to-noise (S/N) $>3$ in the rest-frame UV ($\approx 1500\, \AA $) region of the spectra.

We also include proprietary data for two galaxies from the Mirage or Miracle survey (PIs: P. A. Oesch and R. Naidu; GO-5244); one galaxy (MoM-z11) at $z_{\rm spec}=10.71$ and the previously reported source MoM-z14 at $z_{\rm spec} = 14.44$ \citep{Naidu25}. Complementing the MoM data is a multitude of publicly available data from various \jwst\ programs, including JADES; \#1181, PI: D. Eisenstein, \#1210, PI N. Lützgendorf, \#1286, PI: N. Lützgendorf, \#1287, PI: K. Isaak, \#3215, PI: D. Eisenstein \citep{Eisenstein25}, CEERS; \#1345, PI: S. Finkelstein \citep{Finkelstein23}, \#2750, PI: P. Arrabal Haro \citep{ArrabalHaro23}, CAPERS; \#6368, PI: M. Dickinson, UNCOVER; \#2561, PI: I. Labbe \citep{Bezanson24}, RUBIES; \#4233, PI: A. de Graaff \citep{DeGraaff25}, GLASS; \#3073, PI: M. Castellano \citep{Napolitano25}, DD-2756, PI: W. Chen, GO-1433, PI: D. Coe, and DD-2767, PI: P. Kelly.

There are a total of 48 galaxies in our compiled sample, of which 22 are spectroscopically confirmed to be $z>10$. In the majority of cases (41/48), we are able to identify rest-frame UV or optical nebular emission lines to pin-point the systemic redshift, essential to robustly model the Ly$\alpha$ break and potential damping wings. A broader discussion about the caveats of estimating the redshifts from the Ly$\alpha$ breaks alone for $z>10$ galaxies is provided in a companion paper \citep[Heintz et al. in prep, see also;][]{Hainline24b,Witstok25_ALMA}. The full sample is listed in Table~\ref{tab:1} in Appendix~\ref{app}, along with properties and literature references. 

\section{Analysis} \label{sec:analysis}
\subsection{Emission-line modelling}\label{ssec:lines}

In each spectrum, we identify the most prominent nebular emission lines at the target redshift, ranging from Ly$\alpha$ to the [\oiii]\,$\lambda\lambda 4960,5008$ doublet (up to $z=10$). We modelled each feature with Gaussian line profiles, at vacuum wavelengths, while tying the redshift and intrinsic width. The lines were convolved with the wavelength-dependent Prism resolution, which generally dominates the observed line widths, with an additional factor of $1.3\times$ improvement from the pre-launch predicted resolution \citep[e.g.][]{DeGraaff25}. For the rest-frame UV regime, the following lines were superimposed on the underlying continuum modelled as a single power law, $F_{\lambda} \propto \lambda^{\beta_{UV}}$; \niv]$\lambda1486$, \civ$\lambda\lambda1548,1550$, \heii$\lambda1640$, \oiii]$\lambda\lambda1661,66$, \niii]$\lambda\lambda1746, 1748, 1749, 1752, 1754$, and \ciii]$\lambda\lambda1907,1909$, where the doublets/multiplets are modelled as single Gaussians as they are unresolved in Prism spectra. The inferred UV continuum is also used to model the DLA feature in each galaxy (see Sect.~\ref{ssec:dla} below).

Also of interest are the rest-frame optical emission lines, in particular H$\beta$, which is available in DJA \texttt{v4} NIRSpec spectra for redshifts $z\lesssim10.3$. We measure the strong nebular optical emission lines available for each galaxy, namely [\oiii]$\lambda\lambda4959, 5007$ ($z\lesssim10$), H$\beta$ ($z\lesssim10.3$), H$\gamma$ and [\oiii]$\lambda4363$ ($z\lesssim11.6$), [\neiii]$\lambda3869$ ($z\lesssim13.2$), and [\oii]$\lambda\lambda3727,3729$ ($z\lesssim13.8$). We tie the line fluxes of [\oiii]$\lambda\lambda 4959,5007$ together with a ratio of $1:2.97$ \citep{Storey00}, and model the unresolved [\oii] doublet as a single Gaussian. The underlying continuum for the optical regime is modelled as a simple first-order polynomial. We note that that a majority of the galaxies, especially those in the higher-$z$ range of the sample, have low-S/N spectra and so most emission lines fluxes can be constrained as upper limits only. For galaxies with limited optical lines, $z_{spec}$ is generally recovered from bright UV lines (in all but 7 cases), with the strongest tending to be \ciii]$\lambda1909$ \citep[see also][]{RobertsBorsani25, Tang25, Hayes25}. Where available, we also consider line-redshifts confirmed through auxiliary data, for instance with \alma\ data, e.g. JADES-GS-z11-0 \citep{Witstok25_ALMA}, and JADES-GS-z14-0 \citep{Schouws24, Carniani25}. 

\subsection{Damped Lyman-alpha absorption line modelling}\label{ssec:dla}

To model the Ly$\alpha$ damping wing, we include the full spectroscopic coverage for each galaxy from Ly$\alpha$ to $\lambda \sim 3000\mathrm{\AA}$ rest-frame. This avoids any potential contamination to the fit from the Balmer break at $3645 \mathrm{\AA}$. The intrinsic rest-frame UV continuum is modelled by a single power law $F_\lambda \propto \lambda^{\beta_{UV}}$ as described above. Prominent emission lines from the separate UV slope modelling (Sect.~\ref{ssec:lines}) were added as Gaussian profiles to the intrinsic spectrum. The spectral slope $\beta_{UV}$ was left free in the DLA modelling, as there can be degeneracies between the slope and inferred column density. However, we find the median difference in derived $\beta_{UV}$ between the DLA and line modelling is 0.06 dex, with only 5 objects having a difference of $>0.2$ dex in both, and outside the individual measurement uncertainties. \\ 
We derive the column density $N_{\rm{HI}}$ for each case by approximating the Ly$\alpha$ absorption profile with a Voigt function, following the framework from \cite{TepperGarcia06}. The optical depth is given by:
\begin{equation}
    \tau_{\mathrm{DLA}} = C \: a \: N_{\mathrm{HI}} \: H(a, x)
\end{equation}
where $C$ is the photon absorption constant, $a$ is damping parameter and $H(a, x)$ is the Voigt-Hjerting function. This method has conventionally been used to trace neutral gas from foreground absorbers in the line of sight to bright quasars or gamma-ray bursts, though it has recently been adopted to study gas around singular high-redshift galaxies \citep[e.g.][]{Heintz24_DLA}. This approach assumes the gas dominating the absorption profile is located at a similar redshift to the source, i.e. an early galaxy embedded in a massive neutral gas reservoir, rather than uniformly distributed along the line of sight \citep[though see e.g.][for cases with prominent foreground absorption]{Terp24,Heintz26_natas}. 

\begin{figure*}
    \centering
    \includegraphics[width=1.0\linewidth]{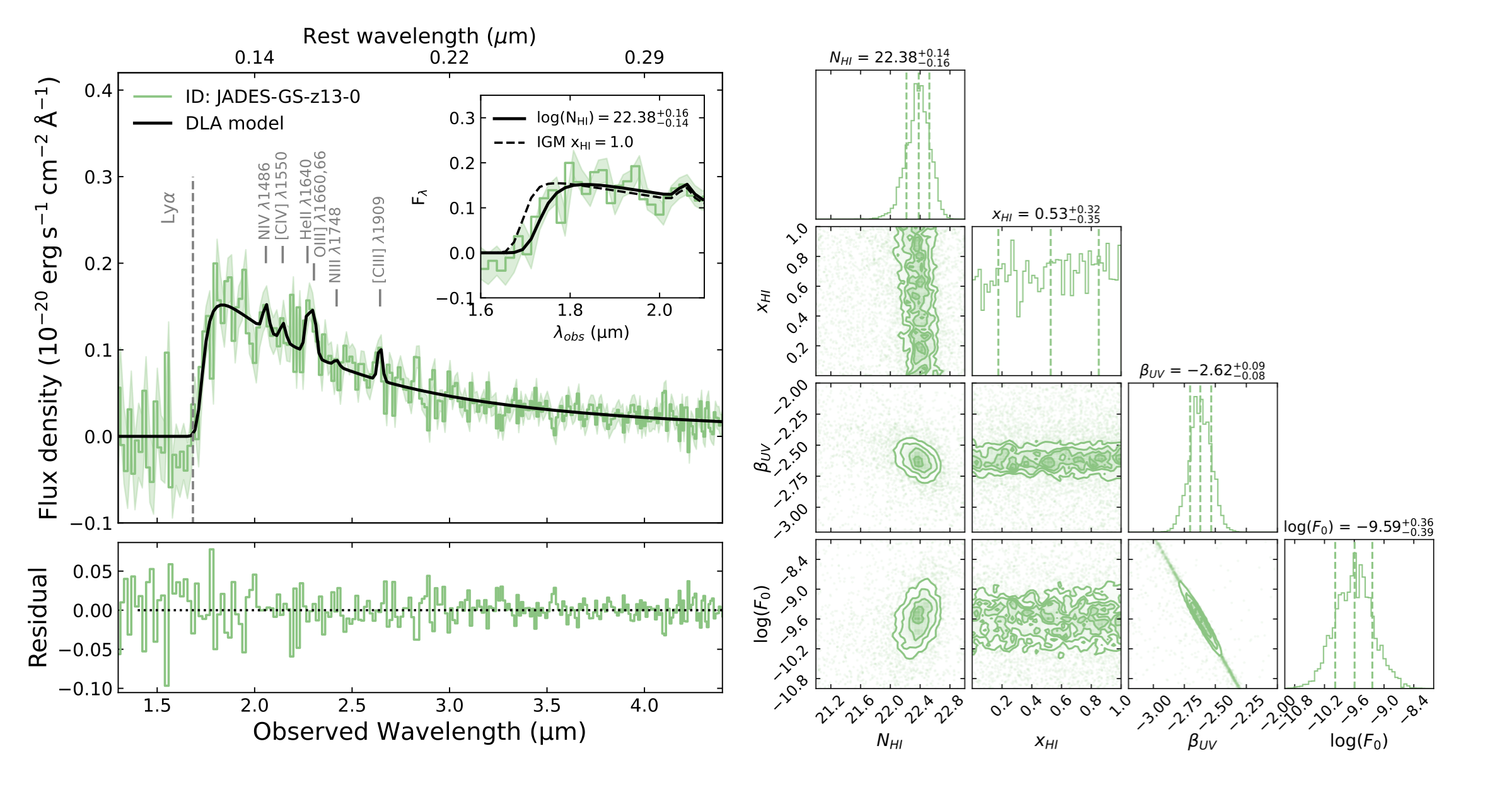}
    \caption{\textit{Left:} Example of UV emission line and DLA fitting for JADES-GS-z13-0 (DJA ID 3215\_20128771) at $z=12.85$. The NIRSpec/Prism spectrum and associated error are shown in green. The marked UV emission lines were modelled and then superimposed on the intrinsic spectrum before modelling the DLA (solid black line). In the inset, we show a zoom on the Ly$\alpha$ region, with the DLA+IGM model as a solid line, and with 100\% neutral IGM only ($\mathrm{x_{\rm HI}}=1.0$) as a dashed line. \textit{Right:} Corner plot of the posterior distributions for the DLA+IGM model, with median, 16th and 84th percentiles marked.}
    \label{fig:DLAex}
\end{figure*}

We also include a prescription for modelling the average fraction of neutral hydrogen in the IGM, as probed in each galaxy sightline. We caution that the neutral hydrogen fraction $x_{\rm HI}$ typically cannot be well constrained due to the low spectral resolution of the Prism configuration \citep{Heintz25, Huberty25, Mason26}, and at large column densities ($\rm N_{\rm HI} \gtrsim 10^{21}\, cm^{-2}$) the damping wing is dominated by the DLA from local \hi\ gas. We adopt the formalism from
\cite{MiraldaEscude98}, to model the absorption profile due to neutral hydrogen in the IGM from the Gunn-Peterson effect:
\begin{equation}
\begin{split}
	\tau_{GP} (z) = 1.8\times 10^5 h^{-1} \Omega_m^{-1/2} \biggl(\frac{\Omega_{b}h^2}{0.02}\biggl) \biggl(\frac{1+z}{7}\biggl)^{3/2} \biggl\langle \frac{N_{\rm HI}}{n_H} \biggl\rangle \\
    \sim 3.88\times10^5 \biggl(\frac{1+z}{7}\biggl)^{3/2}
\end{split}
\end{equation}
We use the correction from \citet{Totani06} to model the contribution to absorption from $x_{\rm HI}$; 
\begin{equation}\label{eq:IGMeq}
\begin{split}
	\tau_{IGM}(\lambda, z_{gal}) = x_{\rm HI} R_\alpha \frac{\tau_{GP(z_{gal})}}{\pi}
\biggl(\frac{1+z_{obs}}{1+z_{gal}}\biggl)^{3/2} \\
\times \biggl[ I\biggl(\frac{1+z_{IGM, u}}{1+z_{obs}}\biggl) - I\biggl(\frac{1+z_{IGM, l}}{1+z_{obs}}\biggl)\biggl]
\end{split}
\end{equation}
Here $R_\alpha = \Lambda_\alpha \lambda_{Ly\alpha}/(4\pi c) = 2.02\times10^{-8}$ is a constant which includes the damping constant of the Lyman-$\alpha$ resonance $\Lambda_\alpha$, $z_{gal}$ is the spectroscopic redshift of the host galaxy, and $z_{obs}$ is the redshift of the neutral gas observed. Equation~\ref{eq:IGMeq} assumes that the neutral hydrogen in the IGM is distributed uniformly between the upper and lower redshift bounds. We set the upper limit $z_{IGM, u} = z_{gal}$, and the lower limit $z_{IGM, l} = 5.3$, when large-scale reionisation is expected to be complete \citep[e.g.][]{Bosman22}. The choice of the upper bound means that the galaxies do not live within ionised regions in the model. For the shape of the damping wing this choice is negligible, and assumes no nebular \lya\ emission. Again, the neutral hydrogen fraction is not well constrained with Prism data, and the choice of lower bound on redshift also does not meaningfully change the derived parameters. 

Combining the intrinsic power law (with emission lines), DLA absorption, and IGM absorption, the total UV spectral range is modelled as:
\begin{equation}
\begin{split}
    F_{\lambda}(\lambda, z_{gal}) = \biggl(F_{0} \lambda^{\beta_{UV}}+\sum^N_i f(\lambda, A_i, \sigma)\biggl)
    \; \times \; \mathrm{exp}(-\tau_{\mathrm{DLA}}(\lambda, N_{\rm HI}, z_{gal})) \\ \times \; \mathrm{exp}(-\tau_{\mathrm{IGM}}(\lambda, x_{\rm HI}, z_{gal}))
\end{split}
\end{equation}
where $\sum^N_i f(\lambda, A_i, \sigma)$ is the sum of all Gaussian line profiles for the identified emission lines. We also include JADES-GS-z13-1-LA \citep[reported by][]{Witstok25_LA} in the sample, where we include an additional Gaussian profile in the model for the \lya\ line, added to the DLA+IGM model. We find a similarly high column density log($N_{\rm HI}/\mathrm{cm}^{-2}$) = $22.86^{+0.20}_{-0.22}$ as reported in \cite{Witstok25_LA} ($\approx 22.8$), albeit from low-S/N spectra, and where the redshift is still uncertain.

An example best-fit model for a more representative high-$z$ galaxy, with prominent UV emission lines and a strong DLA can be seen in Fig.~\ref{fig:DLAex}. The inset of Fig.~\ref{fig:DLAex} highlights the Ly$\alpha$ region and compares two model fits: DLA+IGM absorption with the solid black line, and IGM only (with maximum neutral hydrogen fraction $x_{\rm HI}=1.0$) as the dashed line. The high signal-to-noise of this spectrum clearly constrains that the strong damped UV-turnover cannot be reproduced purely with IGM absorption, and so additional \hi\ gas characterised with a DLA (with $\log (N_{\rm HI}/{\rm cm}^{-2}) = 22.38^{+0.16}_{-0.14}$ for this particular case), is likely present.

For the modelling of the emission lines and the Ly$\alpha$ damping wing, we use {\tt dynesty} \citep{Speagle20} to estimate the posteriors for free parameters and the total evidence of the distribution. 
The priors defined for each parameter for the standard DLA and IGM modelling are $\mathrm{log}_{10}(N_{\rm HI}) \in \mathrm{Uniform}(18, 24)$, $x_{\rm HI} \in \mathrm{Uniform}(0, 1)$, $\beta_{UV} \in \mathrm{Uniform}(-4, 0)$, and $\mathrm{log}_{10}(F_0) \in \mathrm{Uniform}(-17, -4)$. An example of the posterior distribution can be seen in the right panel of Figure~\ref{fig:DLAex}. The best-fit model is convolved with the wavelength-dependent spectral resolution at each step of the optimisation (with the additional factor of $1.3\times$ pre-launch predictions; as with the emission lines).

For objects for which the red wing cannot be decomposed into an IGM and host component (21/48), or damping wings suggesting a negligible DLA contribution, we additionally employ a model with only DLA absorption ($\tau_{\mathrm{DLA}}$) to the intrinsic spectra to obtain upper limits on the total column density (from the 95th percentile of the posterior). This method essentially provides a way to place an upper limit on the damping wing signal from both IGM and DLA absorption, as the two are not discernable in the Prism spectra. We note that generally $x_{\rm HI} = 1.0$ is equivalent to an \hi\ column density of $N_{\rm HI} \approx 10^{21 - 21.5} \mathrm{cm}^{-2}$ at $z=6-14$ \citep{Mason26}. We also test a model with only IGM absorption ($\tau_{\mathrm{IGM}})$ to obtain an upper limit for the neutral hydrogen fraction $x_{\rm HI}$, but the measurement uncertainties are substantial due to the low spectral resolution, and posteriors are generally flat (similarly to the right panel of Figure~\ref{fig:DLAex}), or skewed towards either $0$ or $1.0$. 

The redshifts are typically fixed for DLA fitting, from UV, optical, or \alma\ emission line redshifts described in Section 3.1. As the location of the Lyman break is degenerate with column density, obtaining accurate spectroscopic redshifts are of utmost importance for accurate DLA modelling \cite[e.g.][]{Witstok25_ALMA}. In the sample there are 7 galaxies with ambiguous redshift solutions (mostly $z>10$, see e.g. 6368\_22637), when the S/N of emission line detections are low, and we note that an uncertain redshift may be introducing a larger error in the derived column densities. For these cases we test a model with absorption redshift left free. For the most part, these objects generally have unconstrained column densities in both modelling techniques, and redshifts are consistent within $\pm 0.1$ dex. 


We note also 3215\_20128771 (JADES-GS-z13-0), the example spectrum in Figure~\ref{fig:DLAex}. Our fit results in a 3$\sigma$ detection of both \ciii]$\lambda1909$ and [\oiii]$\lambda\lambda1661,66$, and $z_{spec}=12.85\pm0.02$ which implies a column density of $\mathrm{log}(N_{\rm HI}/{\rm cm^{-2}}) = 22.38^{+0.16}_{-0.14}$. Fixing instead to the potential redshift solution from \citet{Hainline24b} of $z_{spec} = 12.922$, we obtain a virtually identical column density solution of $\mathrm{log}(N_{\rm HI}/cm^{-2}) = 22.36^{+0.14}_{-0.17}$. In a model with free absorption redshift, we find similar solutions though with larger uncertainties (and a tail to lower $N_{\rm HI}$ in the posterior); $z_{abs} = 12.94\pm0.29$ and $\mathrm{log}(N_{\rm HI}/cm^{-2}) = 22.33^{+0.51}_{-2.19}$. 
For these handful of objects, we conclude that potential differences in redshift do not impact the overall findings with respect to the neutral gas content.

We note that \citet{Mason26} highlight the caveat of using an intrinsic power law, rather than creating a continuum model from spectro-photometric modelling of the spectral energy distribution (SED), as power-law fits do not accurately represent the slopes of young stellar populations and nebular continuum emission. For the most extreme UV turnovers predicted in those models with substantial two-photon emission, this may overestimate $N_{\rm HI}$ by up to 1 dex. However, as discussed in Section~\ref{sec:DLAresults}, we find no evidence of dominating nebular continuum emission, and for standard galaxy SEDs with mild to prominent Ly$\alpha$ emission, the uncertainty on the \hi\ column density is only of the order $\lesssim 0.1$\,dex \citep{Heintz24_DLA}. 

\subsection{Rest-frame UV properties}\label{ssec:UV}

To put the derived local \hi\ gas column densities in context, we here aim to characterise the ISM components and star-formation properties of these galaxies; first estimating the dust extinction by calculating $A_V$ from spectro-photometric modelling of the SED. We use the code Bayesian Analysis of Galaxies for Physical Inference and Parameter EStimation \citep[\texttt{BAGPIPES},][]{Carnall2018, Carnall19} to model the SEDs of galaxies using a non-parametric continuity star-formation history \citep[with priors as described in][]{Leja19}. We define a fixed grid of time steps in look-back time where the initial time bin edges are fixed at [0, 3, 10, 25, 50, 100] Myr, after which bin edges are equally spaced in logarithmic look-back time. We allow a flexible dust curve \citep{Salim18}, modified from \cite{Calzetti2000}, which includes a 2175Å UV dust bump $B$ (prior $\mathrm{Uniform(0, 5)}$), and $\delta$ power-law deviation from the \cite{Calzetti2000} attenuation curve slope (prior $\mathrm{Gaussian(\mu=0, \sigma=0.1)}$). An extra factor of $\eta=2$ is applied to the SED-derived $A_V$ values when correcting emission line fluxes, to account for differences between nebular and stellar attenuation \citep{Calzetti2000, Shivaei20, Fisher26}. For the details on the SED-fitting, including star-formation histories and flexible line fitting, we direct the reader to Gottumukkala et. al. (in prep).

\begin{figure*}
    \centering
    \includegraphics[width=0.9\linewidth]{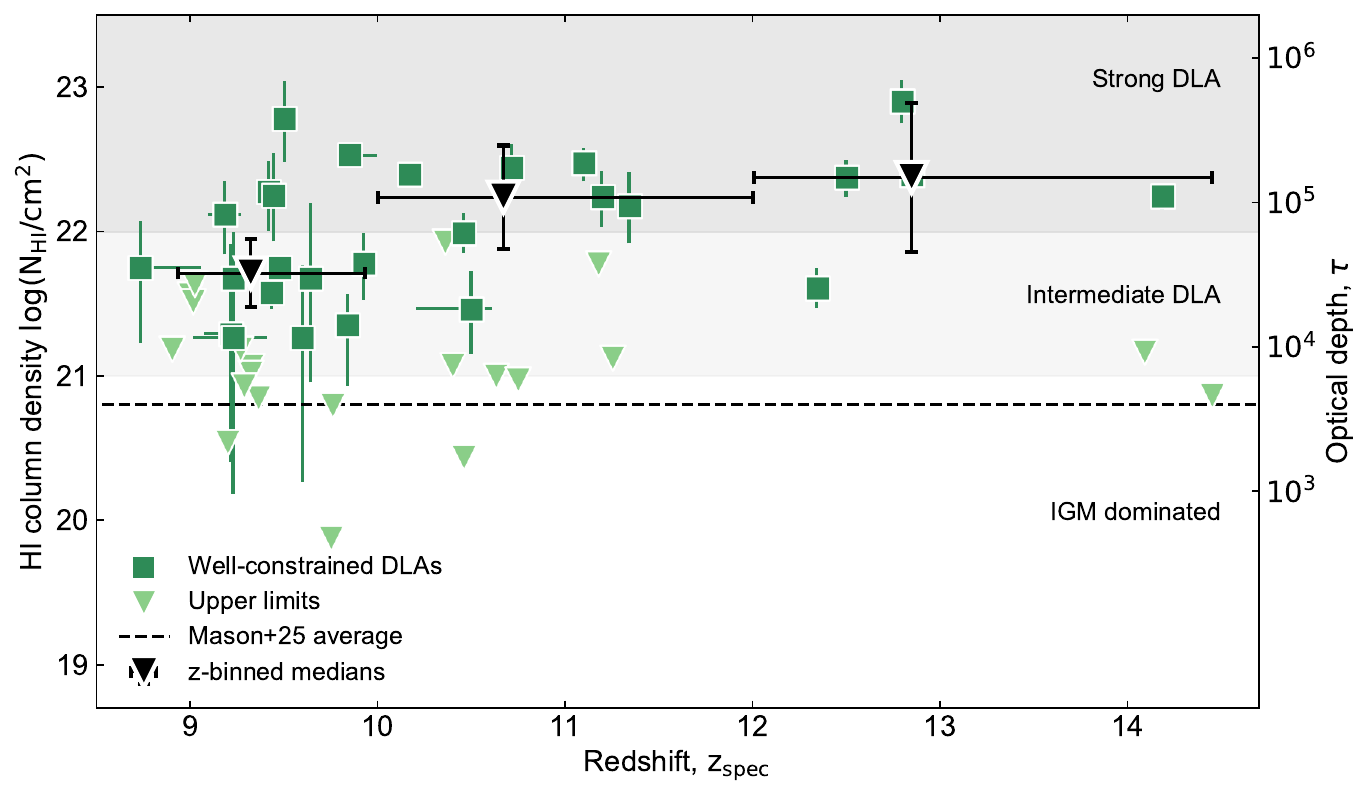}
    \caption{Calculated column densities across the redshift range of the sample, dark green squares are well-constrained column densities with values $N_{\rm HI}>10^{21}\, \rm cm^{-2}$. Upper limits are plotted as triangles, derived from the 95th percentile of the posterior distribution. The different shaded regions represent low column density ($N_{\rm HI}<10^{21}\, \rm cm^{-2}$) where objects are likely IGM-dominated, high column densities ($N_{\rm HI}>10^{21-22}\,\rm cm^{-2}$) where neutral hydrogen exceeds abundance from a fully neutral IGM, and very strong DLAs ($N_{\rm HI}>10^{22}\, \rm cm^{-2}$) with an extreme over-abundance of neutral gas. Regardless of column density, all objects are optically thick with optical depths at the Lyman limit $\tau > 10^3$, with the highest column densities $N_{\rm  HI}=10^{22}\, \mathrm{cm^{-2}}$ corresponding to optical depths $\tau \approx 10^5$. Medians of well-constrained column density binned in redshift ($z<10$, $10<z<12$, $z>12$) are shown as black triangles (with values $N_{\rm HI} = 10^{21.71}, 10^{22.24}$, and $10^{22.37}\, \mathrm{cm^{-2}}$ respectively). These are essentially upper limits, as the fraction of IGM-dominated galaxies/upper limits (which were not included in the median) in each bin is 0.42, 0.50, and 0.29 respectively.}
    \label{fig:z_NHI}
\end{figure*}

The Balmer recombination lines can be used as a tracer of SFR, sensitive to the most massive O-type stars ($M_\ast>10M_\odot$), which contribute significant ionising flux over short lifetimes ($\lesssim 20$Myr). These lines therefore provide an almost instantaneous measure of SFR \citep{Kennicutt98}. However, H$\beta$ is only available for objects $z\lesssim10$, so we rely on UV-based SFRs to remain consistent across the full sample. UV-based SFRs are generally found to be lower, with studies finding typical values of $SFR_{H\alpha}/SFR_{\rm UV} \sim 1-5$ \citep[][Heintz et al. in prep]{Kokorev25,RobertsBorsani25}, though potentially due to spectroscopic samples being biased towards galaxies with rising star-formation histories \citep[e.g.][]{Tacchella22}. 
The overall UV luminosity is also regarded to be a good tracer, sensitive as well to stars of slightly lower masses, and is typically used to trace star formation over longer timescales $\sim100$Myr, although at high-$z$ it may trace substantially lower timescales \citep[e.g. $\sim25$Myr,][]{McClymont25b}. Before calculating the UV magnitude, we first photometrically correct the spectra to account for slit-loss and other reduction issues. Following the process of \cite{RobertsBorsani25}, we rescale the spectra by a constant factor based on the photometric band closest to rest-frame wavelength $1750\mathrm{\AA}$, i.e. F277W for $z>13.5$ and F200W for the remainder of the sample. The UV magnitude at $1500\mathrm{\AA}$ is then determined from the corrected spectrum using a top hat filter of width 100$\textrm{\AA}$ around $1500\AA$ and converted to a luminosity $L_\mathrm{UV}$, with dust, lensing, and K-correction. The SFR is then given by:
\begin{equation}
    {\rm SFR_{UV} (M_\odot yr^{-1}) = 1.0\times10^{-28}} L{\rm _{UV} (erg \,s^{-1} Hz^{-1})}
\end{equation}
This conversion assumes 10\% solar metallicity \citep{Madau14}, and a Salpeter IMF. 

To determine the rest-frame UV sizes of each source, we either adopt the PSF-corrected values reported in individual references (see Table~\ref{tab:1}) or calculate the implied half-light radius, $R_{\rm UV}$, from the measured size in a rest-UV adjacent \jwst\ filter as provided through DJA \citep[e.g.][]{Valentino23}.

\section{Galaxy assembly and star formation at $z>9$} \label{sec:results1}
In this section, we present the main observational results. We focus on the connection between the dense \hi\ gas reservoirs of these early galaxies to their assembly history, star formation, and dust and chemical enrichment. 

\subsection{The \hi\ content of galaxies at $z>9$}\label{sec:DLAresults}

First, we consider the redshift evolution of the \hi\ column density distribution in Figure~\ref{fig:z_NHI}. We find that the median \hi\ column density is $10^{21.71}\, \mathrm{cm^{-2}}$ at $z=9-10$, increasing to $10^{22.24}\, \mathrm{cm^{-2}}$ at $z=10-12$, and $10^{22.37}\, \mathrm{cm^{-2}}$ at $z>12$. The medians should be treated conservatively as upper limits, as they are only calculated from the well-constrained DLA sample. The fraction of upper limits, or IGM-dominated galaxies in each redshift bin are 0.42, 0.50, and 0.29. 

The sample overall displays a mostly bimodal distribution, with the majority of sources having high column densities $N_{\rm HI}>10^{21}\, \rm cm^{-2}$, consistent with probing bulk ISM gas. This limit corresponds to an optical depth at the Lyman limit of $\tau \gtrsim 10^4$, indicating that the gas is fully self-shielding from ionising photons. Sources with column densities below $10^{21}\,\mathrm{cm}^{-2}$ are mostly unconstrained due to the low spectral resolution of the NIRSpec/Prism configuration or difficult to disentangle from pure IGM contribution with $x_{\rm HI} \leq 1.0$.

For these objects we cannot conclusively determine the presence of local \hi\ gas, so instead we mark the upper bounds on the derived column density, taken from the 95th percentile of the $N_{\rm HI}$ posterior distribution. 
The observed \hi\ column density distribution with redshift is largely consistent with previous inferences \citep{Heintz25,Mason26}.  

The broad \hi\ column density distribution indicates complex geometry, implying that we are probing a large variation in sight-lines for the target galaxies, which only span a relatively limited range in $M_{\rm UV}$ and stellar mass. This is consistent with expectations from the zoom-in \textit{SERRA} simulations of galaxies $z\sim6-9.5$, where \citet{Gelli25} find that the variation in column densities measured from a single galaxy can be large (0.5-1.5 dex) due to complex ISM morphologies. 

They also find that despite the fast-changing ISM and clumpy morphology of gas, the bulk of neutral gas is embedded within or close to the central star-forming regions, with minimal contribution from the CGM or filament inflows. We discuss further comparisons to these simulations in Section~\ref{ssec:sims}.

A potential, alternative scenario for the low observed column densities in part of the sources in our sample could be due to an ionised `bubble' \citep{Furlanetto04, Hayes23}. Galaxies with exceptionally powerful ionising photon production may be capable of creating an ionised bubble in their immediate surroundings, as proposed for JADES-GS-z13-1-LA, a \lya\ emitter at $z\approx13$ \citep{Witstok25_LA}, or the most distant current known galaxy MoM-z14 at $z\approx14.44$, which does not display a prominent damping wing, with $x_{\rm HI}$ likely to be $<1.0$ \citep{Naidu25}. 

Additionally, the presence of \lya\ emission could contaminate the observed damping wing, which will not be resolvable with the low-resolution of Prism \citep[e.g.][]{Huberty25}. This has been confirmed with GN-z11, where \lya\ emission is observed in medium-resolution G140M spectra \citep{Bunker23_gnz11}, but undetected in the Prism spectrum, resulting in a derived column density $\mathrm{log}(N_{\rm HI}/{\rm cm^{-2}}) = 19.42^{+1.0}_{-0.87}$, and an upper limit $\mathrm{log}(N_{\rm HI}/{\rm cm^{-2}}) < 21.0$ (95th confidence level). We note, however, that the underlying \hi\ column density could then be intrinsically higher for individual sources, but blended with strong Ly$\alpha$ emission in the Prism spectra. 

\begin{figure}
    \centering
    \includegraphics[width=1.0\linewidth]{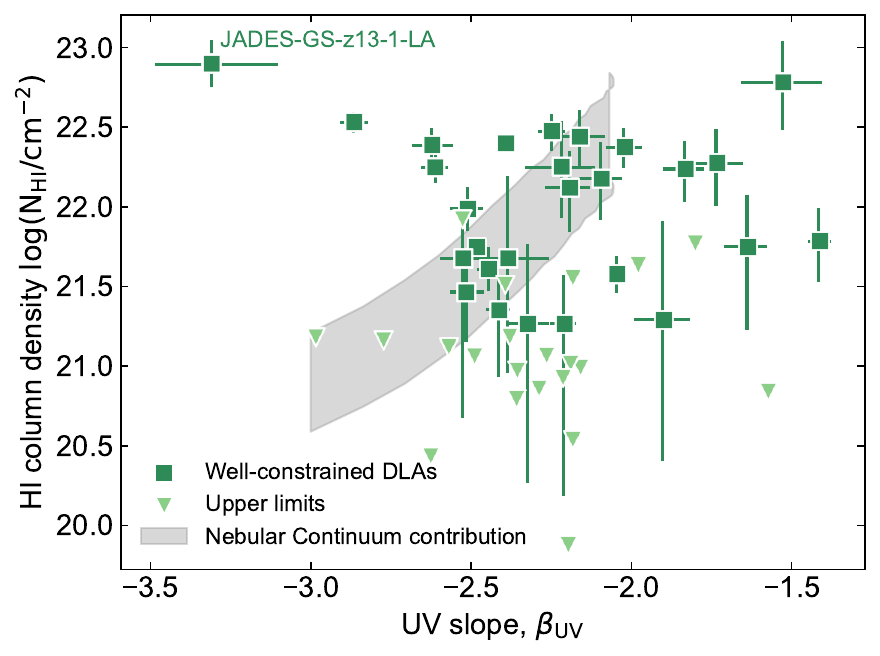}
    \caption{Measured $\beta_{UV}$ slope and column density $\mathrm{log}(N_{\rm HI}/ \rm cm^{-2})$ for the galaxy sample. The green squares and triangles again represent the well-constrained and upper limits for column density respectively, with the grey shaded region representing possible high nebular continuum where two-photon emission may be masquerading as a DLA \citep{Katz25}.}
    \label{fig:beta_NHI}
\end{figure}

Next, we compare the derived \hi\ column densities to the spectral slope, $\beta_{\rm UV}$, in Figure~\ref{fig:beta_NHI}. This is mainly to investigate any potential correlations between the neutral gas column density and the potential dust content or the average age of the stellar population, both increasing with redder $\beta_{UV}$. We do not observe any apparent trend between these two observables (Spearman correlation rank $\rho\approx0.01)$, emphasising again the likely strong sightline-to-sightline variations in $N_{\rm HI}$. As a high nebular continuum contribution may result in two-photon emission being misinterpreted as a DLA \citep[e.g.][]{Cameron24,Katz25,Tacchella25}, we illustrate the region in Figure~\ref{fig:beta_NHI} with a grey-shaded band where this is mostly likely to occur. These values were obtained from \citet{Katz25}, simulating two-photon emission due to stars with temperature $T_{\rm eff}=50,000-100,000$K and density $n_e = 10^3 \rm\, cm^{-3}$ at $z=9$ and measuring the implied observed $\beta_{\rm UV}$ and $N_{\rm HI}$. We have added an additional uncertainty $\pm 0.3$ dex to the inferred masquerading column densities, to account for observational uncertainties and difficulty in fitting the two-photon emission to a DLA model. 


We find ten galaxies with well-constrained \hi\ column densities that lie within the potential nebular continuum region. However, other 
indications of strong nebular continuum emission \citep[see e.g.][]{Trussler25, Cameron24} are inconclusive: \textit{(1)} None of the sample galaxies exhibit prominent Balmer jumps, even when stacked. Although with increasing temperature, the strength of the Balmer jump will decrease. \textit{(2)} Only $4-7$ of 30 galaxies with $z<10$ show strong contribution ($>8\%$) of nebular continuum from H$\beta$ and [\oiii] EWs \citep{Miranda25}, and these do not overlap with the objects within the grey region. \textit{(3)} Similarly, for the 4 galaxies in the full $z>9$ sample with extreme ionising photon efficiency ($\xi_{ion}>10^{26} \mathrm{\; Hz \; erg^{-1}}$), the two populations do not overlap. 

Thus, while we cannot completely rule out the presence of two-photon emission in the sample, we argue that there are no strong indications for this effect, certainly for the majority of the sample. Consequently, we assume in the following analysis that the nebular continuum emission in the target galaxies are negligible, implying also that the estimated column densities are accurate within their measured uncertainties. 

\begin{figure}
    \centering
    \includegraphics[width=1.0\linewidth]{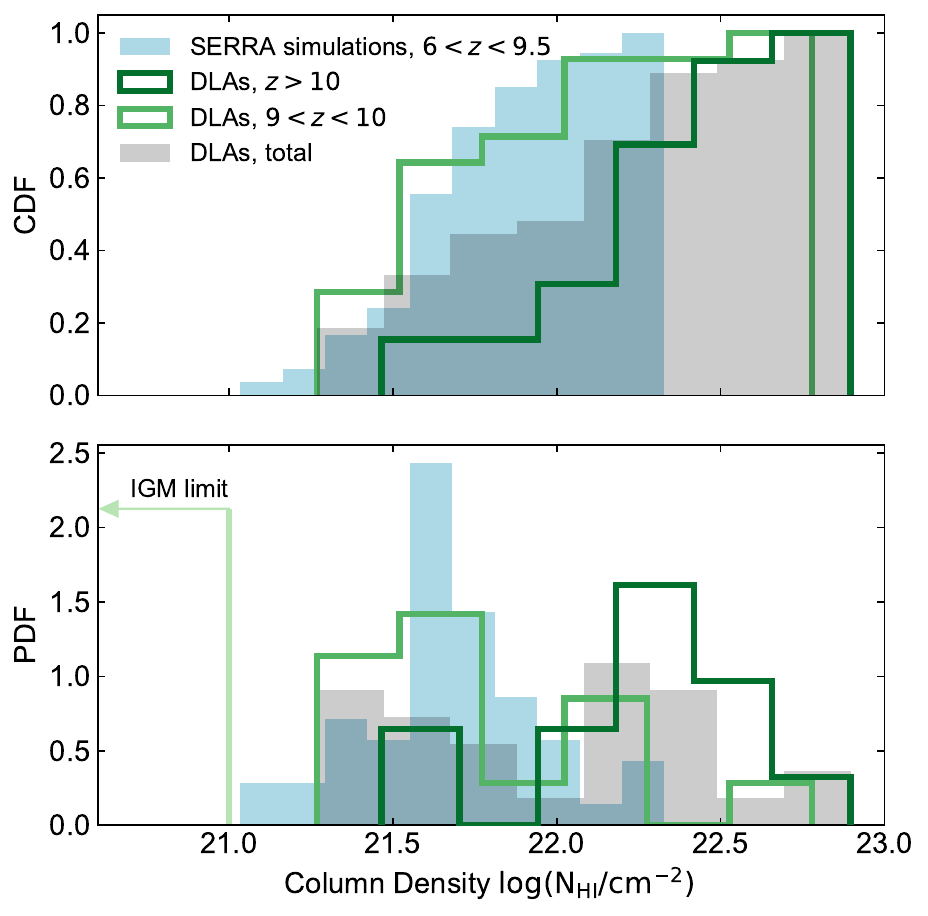}
    \caption{\textit{Top}: Cumulative histogram of derived column densities for our sample in shades of green; separated for non-DLAs (column densities $<10^{21}\,\mathrm{cm^{-2}}$), and the DLA sample (column densities $>10^{21}\,\mathrm{cm^{-2}}$) for $z=9-10$ and $z>10$. The full sample is shown as grey shaded region, comparing directly to the averages of \textit{SERRA}-simulated galaxies $z=6-9.5$ in blue \citep{Gelli25}, showing that while the spread of observed DLAs is broadly consistent, there is a peak of extreme-DLAs ($>10^{22}\,\mathrm{cm^{-2}}$) that are not represented in the averages of simulations, particularly for those $z>10$. \textit{Bottom:} Similar to top panel, but for a normalised PDF, along with the fraction of galaxies with unconstrained DLAs shown as an IGM limit.}
    \label{fig:histogram}
\end{figure}

\subsection{Comparison to simulations}\label{ssec:sims}

We now attempt to place our observations into context of predictions from recent high-resolution zoom-in cosmological simulations. Specifically, we compare our results to the \textit{SERRA} simulations \citep{Pallottini22}, which has enabled a more direct comparison with the \hi\ column densities for galaxies at similar mass and redshift as the sample studied here \citep{Gelli25}. Firstly, we compare the histograms for two distributions in Figure~\ref{fig:histogram}, with the top and bottom panel showing the cumulative and ordinary histograms respectively. The shaded blue region represents the median column density for 50 galaxies in the \textit{SERRA} simulations; averaged over 1000 random sight-lines extending radially from the brightest star-forming region. The grey shaded region shows all DLA-dominated galaxies from observations. The IGM-dominated galaxies with derived column densities $N_{\rm HI}<10^{21}$\,cm$^{-2}$ are represented by the light-green line in the bottom panel.
Although extreme column densities ($N_{\rm HI}<10^{20.5}$\,cm$^{-2}$, $N_{\rm HI}>10^{22.5}$\,cm$^{-2}$) are possible in individual sight-lines in the simulations, they are rare. When averaged, there are no inferred column densities $N_{\rm HI}<10^{21}$\,cm$^{-2}$, and few galaxies showing $N_{\rm HI}>10^{22}$\,cm$^{-2}$ compared to the substantial amount seen in JWST data; especially at $z>10$ (darkest green histogram). This suggests, as also noted by \citet{Gelli25}, that averaging over different pencil-beam sight-lines would be more accurate than how JWST spectroscopy collects light in an aperture, as this would be biased by clumpy distributions. Generally, the observed galaxy spectra will be dominated by emission and absorption from bright star-forming regions. In fact, the percentage of sight-lines which probes column densities $>10^{22}$\,cm$^{-2}$ is similar in both simulations and observations, around 30\%.

\begin{figure}
    \centering
    \includegraphics[width=9.8cm]{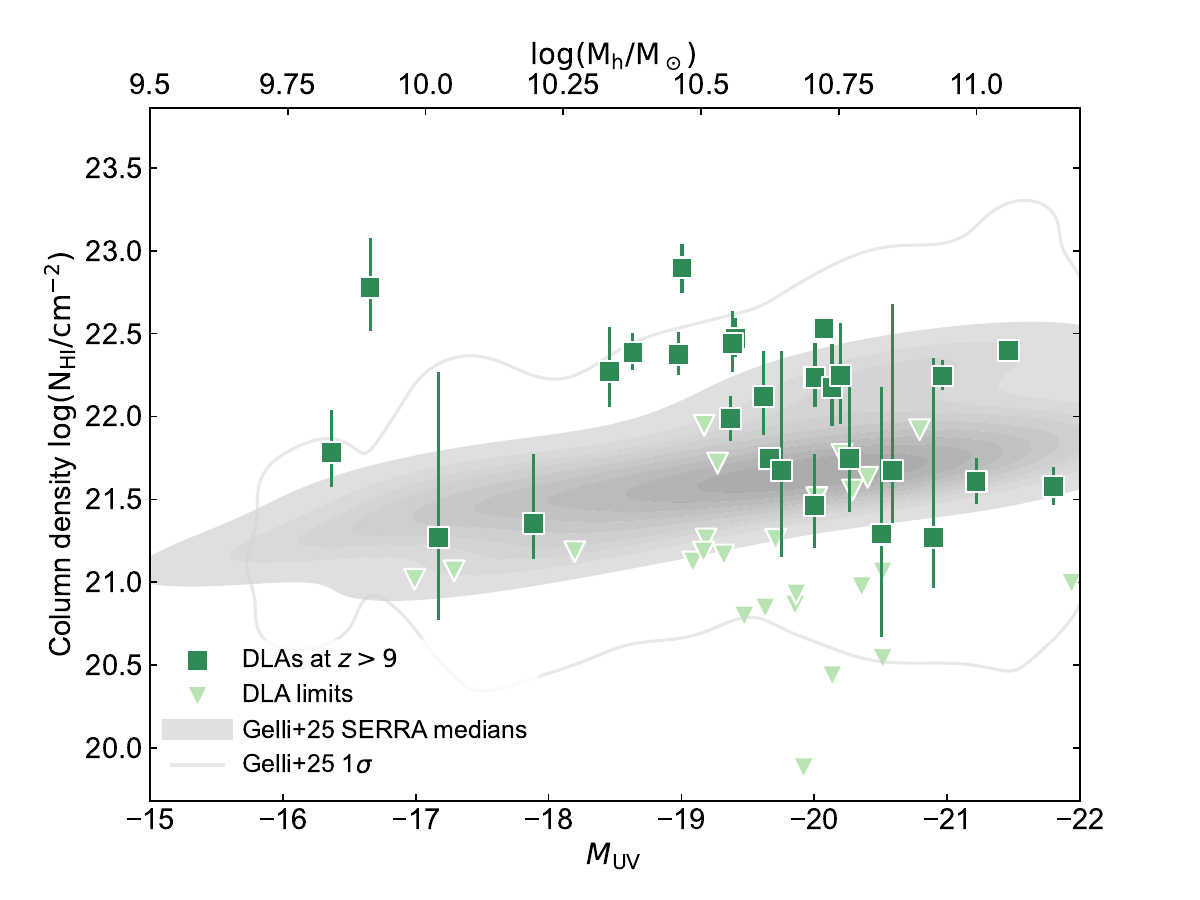}
    \caption{Measured UV magnitude and column density. Halo masses for our objects are derived from UV magnitudes \citep[][no dust model]{Mason23}. The well-constrained column densities of our sample are consistent with results from \textit{SERRA} simulations \citep{Gelli25}. The grey contours represent the average column density of 100 galaxies over random sight-lines, with the 1$\sigma$ error region also shown, representing the scatter across sightlines in simulations.}
    \label{fig:Mhalo}
\end{figure}

In Figure~\ref{fig:Mhalo}, we investigate any potential trends between the derived \hi\ column densities and the observed UV magnitude. As $M_{\rm UV}$ scales with halo mass $M_h$, and we would expect $N_{\rm HI}$ to follow the virial radius trend; e.g. $N_{\rm HI}\sim M_{h}^{1/3}$, we expect a direct correlation. Indeed, the \textit{SERRA} simulations recover a relation $N_{\rm HI}\propto M_h^{0.38}$ for the median $N_{\rm HI}$ for each simulated galaxy, and the scatter of the overall distribution is shown as the filled grey contours in Fig.~\ref{fig:Mhalo}. The unfilled grey contour represents the approximate 1$\sigma$ error region, based on the scatter on $N_{\rm HI}$ for each galaxy. We convert the derived $M_{\rm UV}$ magnitudes to halo masses, $M_h$, following the models from \citet[][assuming no dust]{Mason23}. While we find that the overall scatter in $N_{\rm HI}$ for our observations are consistent with the simulations, we do not recover any strong evolutionary trend with $M_{\rm UV}$. 
If we take the expected local \hi\ column densities from the estimated $M_{\rm UV}$ or halo mass, we can infer the sources that are likely probing excess pristine gas as those significantly above the scatter, with a handful of such cases already visible in Fig.~\ref{fig:Mhalo}.

\subsection{How pristine is the bulk \hi\ gas in galaxies at $z>9$?} \label{ssec:dtg}

There has been significant evidence that ultra-high redshift galaxies ($z\geq8-9$) are dust-poor \citep[e.g.][]{Castellano22, Casey23, Fujimoto23_alma, Carniani25, Bakx25}. 
The so-called `Blue Monsters' \citep{Ziparo23, Ferrara25} with slopes $\beta_{UV}<-2.7$ seen at high-$z$ are evidence for negligible dust-extinction \citep[e.g.][]{Bakx23,Cullen24, RobertsBorsani24, RojasRuiz25}. However, the dust masses seen at $z\sim6$ from \alma\ observations are large \citep[][though see \citealt{Heintz25_a1689}]{Inami22, Fisher25, Watson15, Algera26, Bakx25} and connecting the two populations requires a rapid build-up of dust in $\sim500$~Myr, which is a considerable theoretical challenge. Possible scenarios to explain the dearth of dust at high-$z$ include:
\textit{(i)} grain growth is not yet very effective \citep{Mitsuhashi25}, \textit{(ii)} dust is destroyed or removed \citep{Ferrara23, Ferrara25}, \textit{(iii)} the grain-size distribution is different \citep{Narayanan25,Shivaei25,McKinney25}, resulting in less dust obscuration in the UV. These observations are seemingly at odds with the dense \hi\ gas reservoirs observed here, which require substantial extinction ($A_V \sim 1$\,mag) for the given \hi\ column density and inferred metallicity of the galaxies. 

\begin{figure*}
    \centering
    \includegraphics[width=1.0\linewidth]{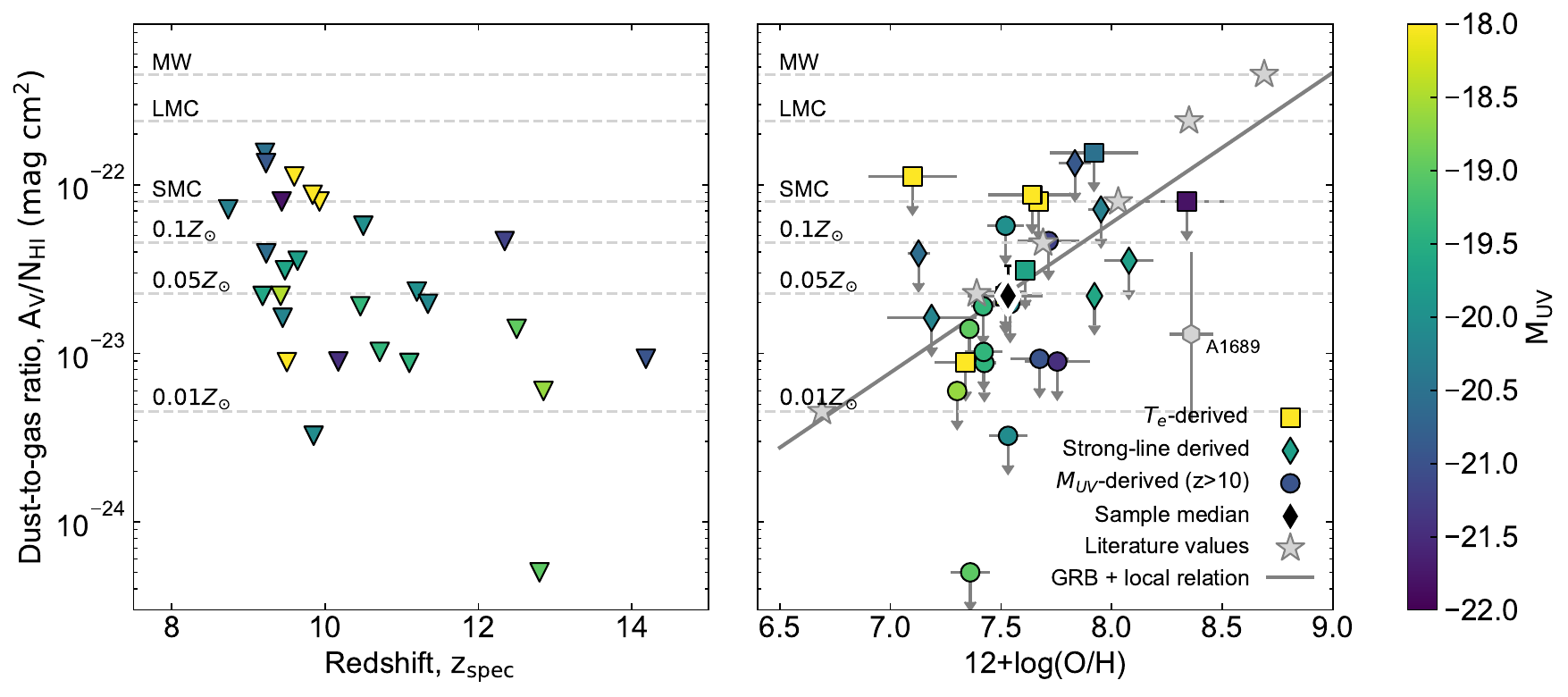}
    \caption{\textit{Left}: Upper limits for the dust-to-gas ratio $A_V/N_{\rm HI}$ against redshift, for the subsample of DLAs with $N_{\rm HI}>10^{21}$\,cm$^{-2}$. Here, $A_V$ are derived from the assumed reddening of the $\beta_{UV}$ slope, and so are taken to be upper limits. The majority of the sample are consistent with being dust-poor, lying below 10\% solar metallicity. \textit{Right}: The trend of $A_V/N_{\rm HI}$ against gas-phase metallicity 12+log(O/H), compared to values from the Milky Way, Large and Small Magellanic Clouds, and 10\%, 5\%, and 1\% metallicities (metallicity for each represented with a star). The black line shows the relation for local galaxies and gamma-ray bursts (GRBs) \citep{Heintz23_GRB}. The method for calculating metallicity for each object is represented by different symbols; each still represents an upper limit in $A_V/N_{\rm HI}$, and is colour-coded with lensing-corrected UV magnitude.} 
    \label{fig:AvNHI}
\end{figure*}

To investigate this in more detail, we here determine the visual extinction $A_V$ for each of the galaxy sight-lines using two approaches; We adopt the $A_V$ derived from the spectro-photometric SED fitting, which overall represents the total reddening of the stellar continuum. Then, we also determine upper bounds on the dust extinction in the line-of-sight by assuming a standard intrinsic slope of $\beta_{UV} = -3$ (expected for the youngest stellar populations and is around the minimum of our sample; not including JADES-GS-z13-1-LA), and that any reddening is due to attenuation according to the steep SMC dust curve. It is not possible to derive the Balmer decrement for the majority of the sources, since H$\beta$ is redshifted out of the NIRSpec coverage for part of the sample at $z\sim10$ and H$\delta$ is normally very faint. Overall, we find that the $\beta_{\rm UV}$-derived $A_V$'s are higher than the SED-derived values by on average 0.12\,mag, as expected if the underlying spectral slope is redder than $-3$ due to a potential older stellar population, or contribution of nebular continuum. 
In the following analysis and results, we use the upper limits on $A_V$ inferred from $\beta_{\rm UV}$, but for any fluxes or properties corrected for dust-extinction we utilise the SED-derived $A_V$ values. 

The dust-to-gas ratio (DTG) is an important property for understanding the ISM composition and the overall chemical enrichment of the target galaxies. Due to its correlation with metallicity \citep[e.g.][]{Zafar13,Heintz23_GRB} it reflects both dust grain growth and chemical evolution. Using the upper limits for dust extinction, we can calculate a proxy for the DTG ratio ($A_V/N_{\rm HI}$) for our sample of well-constrained DLAs, and determine any potential evolution with redshift, see Figure~\ref{fig:AvNHI}. 
For comparison, the average values of $A_V/N_{\rm HI}$ for the Milky Way \citep{Watson11}, and the Large and Small Magellanic Clouds (LMC and SMC) \citep{Gordon03} are shown as well, along with $A_V/N_{\rm HI}$ ratios expected for 10\%, 5\%, and 1\% solar metallicity. The majority of of the sample galaxies at $z>9$ show DTG ratios well below the Milky Way value of $2.2\times10^{-21} \, \mathrm{mag \, cm^{2}}$, with a median DTG ratio of $\sim2.2\times10^{-23}\, \mathrm{mag\, cm^{2}}$, equivalent to $\approx5\%$ solar metallicity according lower-redshift scaling relations \citep[e.g.][]{Heintz23_GRB}. 
While a small number of galaxies have been identified at $z>8$ with metallicities on the order of a few percent solar \citep[e.g.][]{Cullen25}, the majority of galaxies with detectable emission lines show somewhat higher metallicities $Z/Z_\odot \gtrsim10\%$ solar \citep{Pollock25, Hsiao23, AlvarezMarquez25}.

In Figure~\ref{fig:AvNHI}, we also show the DTG against the gas-phase metallicity (oxygen abundance; 12+log(O/H)) for the sample galaxies. The metallicities were calculated either with the direct method for those with high signal-to-noise detections of the [\oiii]$\lambda4363$ auroral line \citep{Pollock25}, strong-line calibrations ($z<10$ galaxies with H$\beta$ detections) from \cite{Sanders24, Laseter2024}, or using the empirical 12+log(O/H)-$M_{\rm UV}$ relation \citep{Pollock25} for those with limited or no detection of UV and optical strong-lines. For comparison we plot the average DTG trend with metallicity for high-redshift GRB sight-lines \citep{Heintz23_GRB}. We find that the DTG ratios derived for our sample generally correlate with the metallicity, noting that the median of our sample lies slightly below the expected relation though still within the $1\sigma$ scatter. We also find a potentially steeper DTG to metallicity relation, with in particular the most metal-poor systems (${\rm 12+log(O/H)} < 8.0$) generally being below the benchmark relation. This could suggest that the absorbing gas is more pristine than the central star-forming ISM \citep[see also e.g.][]{DEugenio24,Heintz25_z14} or that the dust production itself is inefficient. 
We note that most of the metal-poor systems are also those at $z>10$, where 12+log(O/H) was derived using an empirical $M_{\rm UV}$-metallicity relation. It is possible that these objects have $M_{\rm UV}$ which are biased high compared to their true metallicity, which could be expected if they were currently experiencing a strong starburst. If we remove all $z>10$ galaxies from the sample, the median increases from $\mathrm{12+log(O/H)} = 7.53 \rightarrow 7.64$ and $A_V/N_{\rm HI} = 2.2 \rightarrow 3.9 \times 10^{-23}$. These values are within $1\sigma$ of the previous medians, but would reconcile the observations to $\sim$10\% solar as expected. 

For comparison, \cite{Asano13} determine the DTG ratio relative to metallicity for various gas depletion times, showing a steep increase in dust-to-gas mass ratio at a critical metallicity, which is higher if the star formation timescale is shorter. Given the shorter free-fall time expected for more compact high-$z$ objects (see also discussion in Sect.~\ref{ssec:ks} below), it is perhaps unsurprising that the values fall below the local GRB relation.

Further potential evidence for the pristine-gas inflow scenario is the observed offset from the Fundamental Metallicity Relation (FMR, $\rm 12+log(O/H) = log(M_\ast) -\alpha log(SFR)$); a scaling relation constructed to reduce scatter in the mass-metallicity relation (MZR) due to star-formation rate \citep{Mannucci10, Maiolino19, Curti20}. There is a clear deviation from the FMR recovered at high-$z$, constrained both observationally \citep[][]{Heintz23_FMR,Nakajima23,Curti24,Langeroodi23b,Pollock25} and with simulations \citep{McClymont26}. This has generally been attributed to chemical `dilution' of the gas, from pristine gas inflows.
It should be noted that the empirical relations \citep[e.g.][]{Curti20, Andrews13} are extrapolated to the low-mass regime of high-$z$ galaxies, and \cite{Laseter25} show that even at $z=0$, the FMR does not hold for low-mass systems $\mathrm{log}M_\ast<10^9\mathrm{M_\odot}$. Although they find that the high-$z$ galaxies are more offset from the FMR than low-$z$ analogues at a fixed mass, they instead suggest that the offset at high-$z$ is driven by star-formation and enriched outflows. 

\begin{figure}
    \centering
    \includegraphics[width=1.0\linewidth]{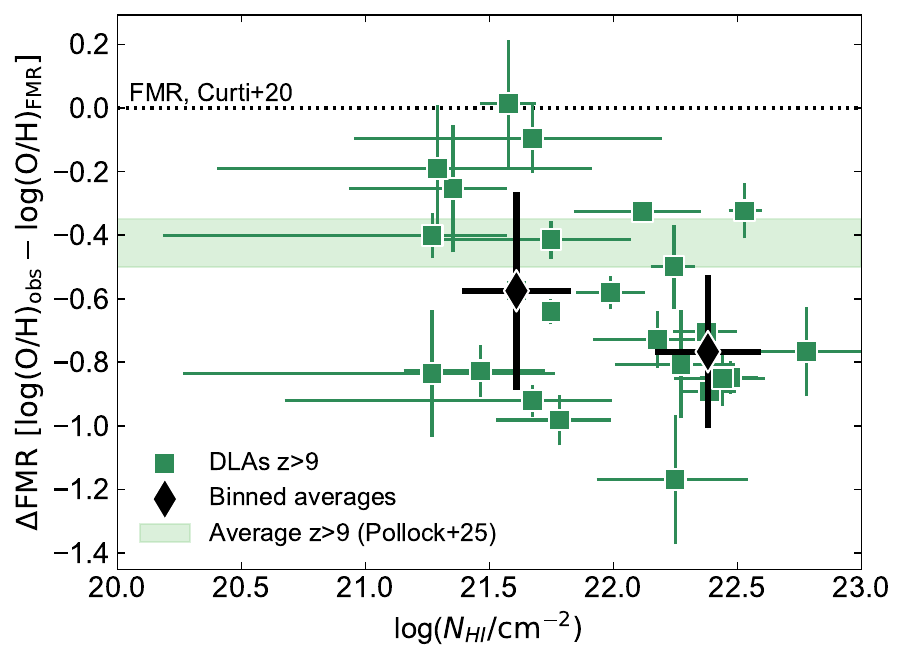}
    \caption{Offset from the Fundamental Metallicity Relation \citep[FMR,][]{Curti20} as a function of the \hi\ column density for the sample of constrained DLAs. Binned averages for $N_{\rm{HI}}=10^{21-22}$\,cm$^{-2}$ and $N_{\rm{HI}}>10^{22}$\,cm$^{-2}$ are shown as black diamonds, with $1\sigma$ error bars. The average offset from metallicity measurements at $z=9-10$ \citep{Pollock25} is shown in light green. There is only a moderate correlation, with Spearman rank $\rho\approx-0.4$, however the strongest DLAs tend to show large offsets from the FMR, as would be expected from pristine gas inflow dilution.}
    \label{fig:FMR}
\end{figure}

To determine whether the abundant \hi\ gas reservoirs probed from the DLAs can explain this potential offset, we compare the physical properties of the sample galaxies at $z>9$ to the local FMR \citep[][with $\alpha=0.56$]{Curti20} to determine FMR offset $\Delta[\rm O/H]$ and show that as a function of \hi\ column density in Figure~\ref{fig:FMR}. We observe only a moderate trend, with a Spearman correlation coefficient $\rho\approx-0.4$. Generally there is a large scatter, but we note that most of the high-column density objects are, on average, more offset from the FMR, even in relation to the median of $z=9-10$ galaxies with metallicities calculated using the [\oiii]$\lambda4363$ auroral line \citep{Pollock25}. We note that the metallicities of galaxies $z>10$ here are calculated using an empirical $M_{\rm UV}$-12+log(O/H) relation. However, excluding these galaxies results in a marginal difference to the binned averages, from $-0.58, -0.77$ to $-0.41, -0.77$, which are well within $1\sigma$. This suggests the galaxies with large neutral gas reservoirs are less enriched at a given SFR-normalised stellar mass. This would support the hypothesis of pristine \hi\ gas diluting the ISM, though a larger sample is needed to robustly validate this trend.

\begin{figure*}[h]
    \centering
    \includegraphics[width=0.8\linewidth]{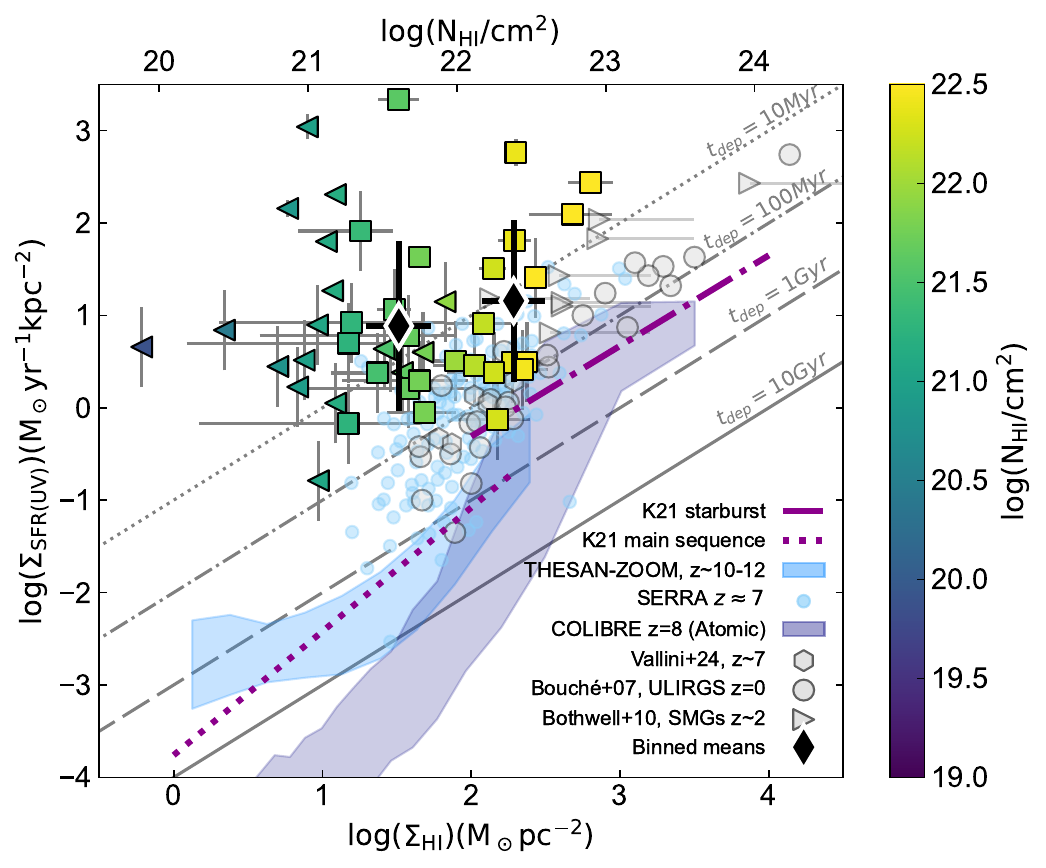}
    \caption{Atomic Kennicutt-Schmidt relation of $\mathrm{log}(\Sigma_{\rm HI})$ and $\mathrm{log}(\Sigma_{\rm SFR})$, coloured with column density values. As previously, we plot non-constrained values as upper limits (95\%). We compare to the canonical \cite{Kennicutt21} (K21) relations for starburst and main sequence galaxies, as well as various gas consumption depletion times corresponding to global star-formation efficiency in grey \citep[][$N=1.0$]{Kennicutt98}. All galaxies lie above the empirical law, with some scatter, and the low column density (non-DLA) objects are extremely offset, even when plotted as upper limits. The binned means for $\Sigma_{\rm gas}=10^1-10^2$, and $10^2-10^3\, \rm M_\odot \, pc^{-2}$ are shown as black diamonds, both consistent with $\sim10\,$Myr. In shades of blue we compare to simulation predictions from \textit{THESAN-ZOOM} at $z=10-12$ \citep{Shen26, Kannan25}, \textit{SERRA} at $z=7$ \citep{Pallottini22}, and \hi-only gas surface density from \textit{COLIBRE} at $z=8$ \citep{Lagos25}. Various literature measurements are shown in grey: from \alma\ detections of [\oiii]\,88$\mu$m and [\cii]\,158$\mu$m \citep{Vallini24} at $z\approx7$,  CO-based measurements for starburst ULIRGs at $z\sim0$ \citep{Bouche07}, and SMGs at $z\sim2$ \citep{Bothwell10, Tacconi08}.}
    \label{fig:KS}
\end{figure*}

\subsection{The Kennicutt-Schmidt relation at $z>9$}\label{ssec:ks}

Observations show that galaxies in the early universe have higher SFRs and SFR surface densities ($\Sigma_{\rm SFR}$) compared to local counterparts \citep[e.g.][]{Heintz23_FMR, Clarke24, Merida25,RobertsBorsani25}, reflecting either intrinsically higher star-formation efficiencies \citep{Fudamoto25, Yung25} or more bursty star-formation histories \citep{Endsley25, Simmonds25}. In this section, we investigate the underlying origin of the luminous galaxies at $z\gtrsim 9$ in context to dense neutral gas reservoirs. Specifically, given that there is now mounting evidence for the DLAs tracing bulk ISM gas, we can directly determine the neutral gas surface density, $\Sigma_{\rm gas}$, integrated over the entire UV-emitting area and compare to $\Sigma_{\rm SFR}$.  

The connection between star-formation and gas surface density was first proposed by \cite{Schmidt59} as a power law $\Sigma_{SFR} \propto (\Sigma_{\rm gas})^N$, and later updated for larger samples of galaxies \citep{Kennicutt98, Kennicutt12, Kennicutt21}, the so-called Kennicutt-Schmidt (KS) law.  

The combination of more compact morphologies and higher SFRs in the early universe implies that high-$z$ galaxies on average tend to have higher SFR surface densities. Consequently, they should lie offset from the canonical KS relation, with higher SFR per unit area. Indeed, \cite{Kennicutt21} include a main sequence and starburst relation, with the latter relation showing both higher SFR and gas surface densities than the main sequence; consistent with shorter gas depletion times, with a median of 240\,Myr compared to 3.2\,Gyr for local spirals \citep{Kennicutt98}. Comparing the derived $\Sigma_{SFR} $ to $\Sigma_{\rm gas}$ for the sample galaxies at $z>9$ thus directly informs the efficiency or typical gas depletion timescales in these young systems. This is particularly crucial, given that the 
KS relation is used in many assumptions in the galaxy formation field; mostly in simulations, and further provides the most direct constraints for the conditions under which the first generation of stars form. 
 

We focus on the UV-derived SFRs, as we can uniformly measure them for the entire sample (H$\beta$ is unavailable $z\gtrsim 10.3$) and adopt the rest-frame UV sizes as derived in Sect.~\ref{sec:analysis}. We define $\Sigma_{\rm SFR}$ as:
\begin{equation}
    \Sigma_{\rm SFR} = \frac{\mathrm{SFR_{\rm UV}}}{2\pi R_{e}^2}
\end{equation}
Where $R_e$ is the half-light or `effective' radius of the galaxy. We caution that for a subset of the sample, the sizes may be overestimated due to being smaller than the point-spread function (PSF) of the \jwst/NIRCam images. This would effectively translate into higher SFR surface densities for the target galaxies.  

As the neutral hydrogen column density $N_{\rm HI}$ represents the total integrated number of hydrogen atoms in the line of sight, it can be converted to an effective surface gas density $\Sigma_{\rm HI}$ by a simple unit conversion: 
\begin{equation}
    {\Sigma_{\rm HI} \, (M_\odot\, {\rm pc}^{-2}) \approx 8\times 10^{-21} \times N_{\rm HI} \, ({\rm HI \: atoms} \: {\rm cm}^{-2})}
\end{equation}
This conversion assumes that the derived column density represents the average volumetric density of the \hi\ gas and that the line of sight is representative of the average across the whole galaxy (which might show large variations, as discussed in Sect.~\ref{ssec:sims}). 

While the canonical KS relation describes the \textbf{total} gas surface density $\Sigma_{\rm gas} \equiv \Sigma_{\rm HI} + \Sigma_{\rm H_2}$, our measurements trace the neutral, \textit{atomic} gas only. In the following, we assume that the bulk of gas present in the $z>9$ galaxies is neutral. 
Since molecular hydrogen $\mathrm{H_2}$ typically forms on interstellar dust grains \citep{Cazaux09, Wakelam17}, its production will likely be inefficient in the observed low-metallicity, dust-poor environments. Intense UV-photons (Lyman-Werner feedback) will also increasingly photo-dissociate $\mathrm{H_2}$ \citep{Nebrin23, Sugimura24}, especially in low-metallicity environments. Additionally, it has been suggested that star-formation can occur directly from atomic hydrogen, without a molecular gas phase if the metallicity is low (on the order of a few percent) \citep{Krumholz12}.
Although we are not claiming that molecular gas in these systems is completely absent, given the age of the universe and the relatively low metallicities measured, we are approaching a regime where it is possible that recent Population III star formation may have dominated. 
The assumption that the majority of gas present in the high-redshift galaxy population in its atomic, neutral form is thus physically motivated. 


In Figure~\ref{fig:KS}, we compare the derived $\Sigma_{\rm gas}$ and $\Sigma_{SFR}$. 
As in the previous analysis, the `unconstrained' galaxies, with Ly$\alpha$ damping wings consistent with IGM absorption only, are plotted as 95th percentile upper limits. Interestingly, the IGM-dominated objects do not follow the expected trend of the KS law, with similar, or even higher SFR surface densities than the sources with strong DLAs. We plot the means of $\Sigma_{\rm gas}$ bins ($10^1-10^2$, and $10^2-10^3\, \rm M_\odot \, pc^{-2}$ respectively), both of which lie around the $10\,$Myr depletion timescale. 

The empirical KS relations are overlaid for comparison, showing the main sequence and starburst regimes \citep{Kennicutt21}, as well as the canonical KS law with different gas depletion times $0.1, 1$, and $10$\,Gyr. Assuming a free-fall time ($t_{ff}$) of $10^8$ years, typical for a Milky Way type galaxy, these depletion times would correspond to star formation efficiencies (SFE) $\epsilon = 100\%, 10\%, 1\%$ respectively \citep{Kennicutt98}. However, as free-fall time scales with $1/\sqrt{\rho}$; denser and more compact high-$z$ galaxies would consequently have a much lower $t_{ff}$, on the order of 1Myr for $\Sigma_{\rm gas} \approx 10^3 \mathrm{\: M_\odot \, pc^{-2}}$. 

Simulation predictions for total $\Sigma_{\rm{HI+H_2}}$ from \textit{THESAN-ZOOM} at $z=10-12$ \citep[][light blue shaded region]{Kannan25, Shen26}, and \textit{SERRA} at $z\approx7.7$ \citep[][light blue scatter points]{Pallottini22} are shown in Figure~\ref{fig:KS} for comparison. 
The \textit{SERRA} galaxies have high star-formation rate surface densities due to high SFR and small UV sizes, and are therefore also located above the local KS law. \cite{Pallottini22} suggest this result is due to burstiness, with \textit{SERRA} galaxies on average $3\times$ more bursty than galaxies lying on the KS relation for the same gas surface density (i.e. located above the Kennicutt-Schmidt relation by a factor $\kappa_s = 3$). 
We also show atomic-only $\Sigma_{\rm{HI}}$ from the \textit{COLIBRE} simulations at $z=8$ \citep{Lagos25} in the dark blue shaded region, which show a similar range of gas densities though lower star-formation rate densities; possibly due to a fixed efficiency per free-fall time of $\epsilon_{ff} = 1\%$.
Though there are galaxies in the \textit{COLIBRE} simulations with \hi\ depletion times of $\sim100\,$Myr, they tend to be higher gas metallicity (though still sub-solar by $0.3-0.5$ dex) and high $\Sigma_{\rm SFR}$ objects. Shorter depletion times of $10-100\,$Myr are seen in \textit{COLIBRE}, though only for $H_2$ depletion time. 

Finally, to put our results in context with galaxies across cosmic time, we consider various literature values across a wide redshift range including: Gas densities derived from \alma\ detections of [\oiii]88$\mu$m and [\cii]158$\mu$m from \cite{Vallini24} at $z\approx7$, and CO-based measurements for starburst ULIRGs at $z\sim0$ \citep{Bouche07}, and SMGs at $z\sim2$ \citep{Bothwell10, Tacconi08}. 
While these literature measurements are generally above the typical KS relation, with implied depletion times $t_{dep}\sim 100$Myr, the DLA measurements for the sample galaxies at $z>9$ are still systematically higher, indicating a likely transition in the conditions under which stars in galaxies at $z>9$ form. Reconciling the observations with the canonical KS relation would imply an additional molecular or `hidden' gas reservoir $300\times$ more massive than the neutral gas present, which we consider unlikely, though note that some of the discrepancy could arise from \lya\ emission contamination as described previously.

The  star-formation rate efficiency is defined as $\epsilon = t_{\rm ff}/t_{\rm dep}$, where $t_{\rm ff}$ is the free-fall time of the gas cloud, $t_{ff} = (3\pi/32G\bar{\rho})^{1/2}$ and the depletion time defined as $t_{dep} \equiv \Sigma_{\rm gas}/\Sigma_{SFR}$ \citep{Kennicutt98}. For the sample galaxies with strong DLAs at $z>9$, the mean free-fall time and depletion timescales are $t_{\rm ff}\approx25$Myr (corresponding to a particle density of $n\sim0.03 \, \rm cm^{-3}$) and $t_{\rm dep}\sim28$Myr, respectively. 
This suggests a mean efficiency $\epsilon=90\%$. This is extremely high, approaching the 100\% efficiencies that are claimed to be necessary to explain the UV-bright population. Near-unity efficiencies have been produced in simulations with feedback-free systems \citep{Somerville25, Dekel23, BoylanKolchin25}, though it is currently debated whether feedback-free can exist in systems due to \lya\ feedback \citep{Manzoni25, Ferrara25_Lya, Nebrin25}. In this framework, 100\% efficiency would require extreme surface densities $\sim10^5\,M_\odot \, {\rm pc}^{-2}$ and near-solar metallicities. In summary, our results strongly imply efficient conversion of gas into stars at high redshifts, either quantified as a rapid free-fall time due to the higher gas densities, $t_{\rm ff}\propto 1/\sqrt{n_{\rm H}}$, or the standard efficiency parameter reaching $\epsilon \approx 90\%$.

\subsection{Caveats}
With an average efficiency of the sample reaching 90\% and many individual galaxies exceeding unity efficiencies, we now summarise any possible biases, systematics, or caveats with our assumptions that may affect the interpretation of Figure 8. 

\noindent $\bullet$ Firstly, we measure UV SFRs over an assumed $10^8$ year timescale. If the star-formation is especially stochastic, or bursty \citep[e.g.][]{Sun23} then the SFR we measure may not be representative of the system across $10^8$ years, biasing us to higher $\Sigma_{\rm SFR}$. However, at high-$z$ it has been suggested that the timescales would be closer to $25\rm \, Myr$ \citep{McClymont25b}. Should the effective timescale even decrease to $10^7-10^6$ years, the offset in $\Sigma_{\rm SFR}$ will be $\sim 0.2-1$ dex lower \citep{Madau14}. However, we note that other measures of SFR, such as using H$\beta$ luminosity (when available), [\ciii] luminosity (Heintz et al. in prep), or SED-derived values on 10 or 100 Myr timescales result in similar, or higher SFR surface densities. \\
$\bullet$ Similarly, if the IMF is top-heavy, the true SFRs would be lower than the measured values \citep[e.g.][]{Hutter25}.\\
$\bullet$ Accurately measuring the effective radii of the galaxies is perhaps the largest uncertainty in the work, as many of the galaxies are extremely compact and close to the resolution limit. If the full sample had physical sizes $\sim10\times$ larger, the mean log($\Sigma_{\rm SFR}$) would be $\sim2$ dex lower, more in line with the starburst regime from \citet{Kennicutt21}. 
However, we note that for many of the galaxies, the true size is more likely to be smaller than the measured size (due to being smaller than the resolution limit), resulting in higher $\Sigma_{SFR}$. \\
$\bullet$ With the PRISM spectra we could be missing contamination from Lyman-alpha emission blended with a damping wing. This could result in an apparent steep drop-off (i.e. like GN-z11’s Prism spectrum) which would result in a lower constrained column density. This depends on a non-zero covering fraction, but may be biasing particularly the ‘unconstrained DLAs’ to lower gas surface densities. \\
$\bullet$ The large observed range in the column densities are likely biased by the geometry of these systems. We are assuming a screen-like distribution when converting the column density to surface gas density, assuming the distribution of gas is roughly the size of the UV region (i.e. with a covering fraction $f_c = 1$). If there are high column densities of gas extending outwith this region \citep[i.e. suggested by ][]{Rowland25_DLA} our gas densities are biased too low. 
Additionally, if the gas is instead clumpy or patchy, the derived $N_{\rm HI}$ should be interpreted as a UV-luminosity-weighted column density, and the true area-averaged $\Sigma_{\rm gas}$ would be larger according to the covering fraction, $\sim 1/f_c$.\\
$\bullet$ Due to selection effects, we are more likely to observe highly star-forming galaxies at high-$z$, which must require large gas supplies, potentially biasing our sample towards objects with higher $N_{\rm HI}$. \\
$\bullet$ As mentioned previously, we would need on average $\sim300\times$ more hidden or molecular gas for the current observations to lie on the starburst \citet{Kennicutt21} relation. There indeed must be some molecular gas present in these systems, though we believe not on the order of $100-1000\times$ the inferred neutral gas. \\

While it is unlikely that any of these effects alone are able to explain the discrepancy, it is possible that invoking a few simultaneously could decrease the offset. Taking into consideration each of the caveats listed above, it is possible that the current observed tension with the local Kennicutt-Schmidt relation is not due to low-dust and increased star-formation efficiency, but a combination of bursty star-formation, top-heavy IMF, larger UV sizes, low covering fractions, or exceptionally high molecular gas content.

\section{Summary and Future Outlook}\label{sec:conclusion}

In this work, we have performed a careful analysis of the \lya\ damping wings observed in the total compiled sample of 48 UV-bright galaxies at $z>9$, near the expected onset of cosmic reionisation. All galaxies have been observed with \jwst/NIRSpec in the Prism configuration from various observing programs, and reduced uniformly through the optimised DJA framework. 

The main goal was to measure the local \hi\ gas producing strong damping wings. This was crucial to investigate the \hi\ gas mass build-up in these early galaxies, and determine its role in governing the observed chemical enrichment, dust properties, and star-formation rate densities.

We then compared the neutral gas surface densities $\Sigma_{\rm gas}$ integrated over the UV-emitting region as probed via $N_{\rm HI}$ to the effective SFR surface density, $\Sigma_{\rm SFR}$, to gauge the efficiency of the observed star formation. \\~\\
Our main findings are summarised below:\\
$\bullet$ We found that the \hi\ column density distribution of DLAs appeared to evolve with redshift, with median $N_{\rm HI} = 10^{21.71}$, $10^{22.24}$, and $10^{22.34} \, \mathrm{cm^{-2}}$ at $z<10$, $10<z<12$, and $z>12$, respectively. The majority of the sample galaxies show prominent DLAs, reaching column densities $N_{\rm HI}\gtrsim 10^{22.5}\,{\rm cm^{-2}}$. We highlight the minority subsample of `DLA-free' objects at $z>9$, which are of potential interest for \lya\ follow-up surveys with higher resolution spectra, since the steepness of the \lya\ wings might reflect \lya\ emission that is unresolved in the Prism resolution. \\
$\bullet$ The overall \hi\ column density distribution was found to be in good agreement with predictions from simulated galaxies at $z\gtrsim 8$ from the \textit{SERRA} suite of zoom-in high-resolution cosmological simulations, also considering their absolute magnitudes $M_{\rm UV}$ and inferred halo masses, though with a larger fraction of high column density systems at $z>10$. This indicated that the gas probed by DLAs are likely already driving the central star formation in the targeted galaxies.\\
$\bullet$ We found that the dust-to-gas ratios, $A_V/N_{\rm HI}$, were in good agreement with lower-redshift empirical relations based on their metallicities, though we noted a potential decrease in $A_V/N_{\rm HI}$ in the most metal-poor systems, with ${\rm 12+log(O/H)} < 8.0$. Higher resolution spectroscopy is, however, needed to robustly constrain the metal content of the foreground DLA gas via low-ionisation metal lines as commonly done for GRB or quasar absorption line systems. \\
$\bullet$ We further tested the proposed scenario of the abundant neutral gas causing the deviation in the fundamental-metallicity relation towards lower metallicities by comparing the offset directly to the derived \hi\ column densities. We indeed found a trend for the sources with the lowest metallicities at a given stellar mass and SFR to show the highest $N_{\rm HI}$ at $z>9$, though still with a substantial scatter. \\
$\bullet$ We found that all the sample galaxies at $z>9$ were offset from the canonical KS relation, with most having depletion times well below 100\,Myr, implying highly efficient and rapid star formation from the available gas. The derived gas surface densities were found to be consistent with predictions from state-of-the-art cosmological simulations, though none of them reaching the SFR surface densities observed here. These deviations from the local KS relation and the more rapid gas depletion times are critical to incorporate into future simulations to understand the conditions under which the first stars and galaxies form. \\
For future avenues, we strongly emphasise the need higher-resolution \jwst\ grating spectroscopy to better constrain the presence of \lya\ emission, the \lya\ damping wings to disentangle \hi\ in the IGM from local galaxy contributions, and measure the metallicities of the absorbing gas from low-ion metal absorption lines directly. This will greatly advance our understanding of the early assembly and formation of stars and galaxies, and their impact on the large-scale reionisation of the Universe.  

\begin{acknowledgements}
The data products presented herein were retrieved from the Dawn JWST Archive (DJA). DJA is an initiative of the Cosmic Dawn Center, which is funded by the Danish National Research Foundation under grant DNRF140.
We express our greatest gratitude to the investigators on the major JWST observing programs, such as RUBIES, CEERS, CAPERS, UNCOVER, and JADES. The work presented here would not have been possible without their major efforts in designing and obtaining the observational data included in our work here. 
KEH acknowledges support from the Independent Research Fund Denmark (DFF) under grant 5251-00009B and co-funding by the European Union (ERC, HEAVYMETAL, 101071865). Views and opinions expressed are, however, those of the authors only and do not necessarily reflect those of the European Union or the European Research Council. Neither the European Union nor the granting authority can be held responsible for them. 
P. Dayal warmly acknowledges support from an NSERC discovery grant (RGPIN-2025-06182). 
JRW acknowledges that support for this work was provided by The Brinson Foundation through a Brinson Prize Fellowship grant. 
This work is based in part on observations made with the NASA/ESA/CSA James Webb Space Telescope. The data were obtained from the Mikulski Archive for Space Telescopes (MAST) at the Space Telescope Science Institute, which is operated by the Association of Universities for Research in Astronomy, Inc., under NASA contract NAS 5-03127 for JWST. This work has received funding from the Swiss State Secretariat for Education, Research and Innovation (SERI) under contract number MB22.00072, as well as from the Swiss National Science Foundation (SNSF) through project grant 200020\_207349. 
Software used in this work includes \textsc{matplotlib} \citep{Matplotlib}, \textsc{numpy} \citep{Numpy}, \textsc{astropy} \citep{Astropy}, \textsc{scipy} \citep{scipy}, \textsc{pandas} \citep{pandas}, and \textsc{pyneb} \citep{Luridiana15}. 
\end{acknowledgements}

\bibliography{ref}{}
\bibliographystyle{aasjournal}

\onecolumn
\appendix
\section{Table and Spectra of all galaxies}\label{app}

\renewcommand*{\arraystretch}{1.4}
\setlength\LTleft{-1.0cm}
\begin{longtable}{c|c|c|c|c|c|c}
\caption{ID, spectroscopic redshift, UV magnitude, UV slope, \hi\ column density, and literature references for the full $z>9$ sample}\label{tab:1}\\
\hline
ID & Alternative name & z$_{\rm spec}$ & $M_{\rm UV}$ &
$\beta_{\rm UV}$ & log($N_{\rm HI}/\mathrm{cm}^{-2}$) & References\\
\hline\hline
\endfirsthead

\caption{continued.}\\
\hline\hline
\endhead
    $5224\_277193$ & MoM-z14 & $14.44^{+0.01}_{-0.02}$ & $-19.86 \pm 0.15$ & $-2.32 \pm 0.15$    & $19.42^{+0.98}_{-1.01}$ & (1) \\
    $1287\_183348^{(*)}$ & JADES-GS-z14-0 & $14.18^{+0.03}_{-0.03}$ & $-20.96 \pm 0.09$ & $-2.53 \pm 0.06$ & $22.25^{+0.13}_{-0.11}$ & (2, 3, 4, 5, 6, 7, 8) \\
    $1287\_20018044$ & JADES-GS-z14-1 & $14.08^{+0.16}_{-0.03}$ & $-19.32 \pm 0.08$            & $-2.73 \pm 0.14$ & $19.54^{+1.08}_{-1.08}$ & (5, 9) \\
    $1287\_20013731$ & JADES-GS-z13-1-LA & $13.0^{+0.2}_{-0.2}$ & $-19.00 \pm 0.40$            & $-3.04 \pm 0.42$ & $22.86^{+0.22}_{-0.20}$ & (10, 11, 12) \\
    $3215\_20128771$ & JADES-GS-z13-0 & $12.85^{+0.02}_{-0.02}$ & $-18.63 \pm 0.04$            & $-2.6 \pm 0.08$ & $22.38^{+0.16}_{-0.14}$ & (13, 14, 15, 16) \\
    $3215\_20096216$ & JADES-GS-z12-0 & $12.5^{+0.01}_{-0.01}$ & $-18.98 \pm 0.04$ & $-1.97 \pm 0.08$ & $22.36^{+0.19}_{-0.16}$ & (14, 15, 17) \\
    $3073\_22600^{(*)}$ & GHZ2/GLASS-z12 & $12.34^{+0.01}_{-0.01}$ & $-21.22 \pm 0.04$ & $-2.45 \pm 0.06$ & $21.60^{+0.21}_{-0.18}$ & (18, 19, 20, 21, 22) \\
    $2750\_10$ & CEERS-10 & $11.2^{+0.06}_{-0.05}$  & $-20.00 \pm 0.11$ & $-1.76 \pm 0.1$       & $22.21^{+0.32}_{-0.24}$ & (23, 24, 25) \\
    $6368\_126973$ & CAPERS-126973 & $11.17^{+0.10}_{-0.05}$ & $-20.21 \pm 0.16$ & $-2.1 \pm 0.13$ & $19.87^{+1.33}_{-1.43}$ & (26) \\
    $2750\_1$ & Maisie's/CEERS-16943 & $11.34^{+0.02}_{-0.02}$ & $-20.13 \pm 0.06$ & $-2.11 \pm 0.1$ & $22.12^{+0.52}_{-0.31}$ & (27, 28, 29, 23, 24) \\
    $1210\_14220$ & JADES-GS-z11-0 & $11.1^{+0.03}_{-0.04}$  & $-19.40 \pm 0.04$ & $-2.22 \pm 0.06$ & $22.47^{+0.17}_{-0.15}$ & (30, 14, 15, 16) \\
    $1287\_20015720$ & GS-z11-1 & $11.25^{+0.02}_{-0.02}$ & $-19.09 \pm 0.08$ & $-2.64 \pm 0.08$ & $19.47^{+0.96}_{-1.09}$ & (6) \\
    $1181\_3991$ & GN-z11 & $10.63^{+0.01}_{-0.01}$ & $-21.94 \pm 0.01$ & $-2.22 \pm 0.02$      & $19.42^{+1.00}_{-0.87}$ & (31, 32, 33, 34, 35, 36) \\
    $6368\_22637$ & CAPERS-22637 & $10.75^{+0.15}_{-0.09}$ & $-20.36 \pm 0.09$ & $-2.43 \pm 0.09$ & $19.48^{+1.03}_{-1.08}$ &  \\
    $5224\_154407$ & MoM-z11-2 & $10.71^{+0.04}_{-0.06}$ & $-19.38 \pm 0.17$ & $-2.16 \pm 0.11$ & $22.41^{+0.32}_{-0.24}$ & (26) \\
    $6368\_136645$ & CAPERS-136645 & $10.5^{+0.29}_{-0.11}$ & $-20.00 \pm 0.05$ & $-2.55 \pm 0.08$ & $21.38^{+0.55}_{-0.37}$ & (26) \\
    $1287\_20176151$ & JADES-GS-20176151 & $10.46^{+0.11}_{-0.04}$ & $-19.37 \pm 0.06$ & $-2.49 \pm 0.08$ & $21.98^{+0.20}_{-0.18}$ &  \\
    $3073\_22302^{(*)}$ & GHZ7& $10.46^{+0.03}_{-0.03}$ & $-20.14 \pm 0.05$ & $-2.64 \pm 0.07$ & $19.24^{+0.83}_{-0.93}$ & (37) \\
    $3073\_23984^{(*)}$ & GHZ8 & $10.36^{+0.10}_{-0.14}$ & $-20.79 \pm 0.09$ & $-2.49 \pm 0.11$ & $20.51^{+1.66}_{-0.93}$ & (37)  \\
    $1433\_3349^{(*)}$ & JD2 & $10.17^{+0.02}_{-0.02}$ & $-21.46 \pm 0.05$ & $-2.34 \pm 0.03$   & $22.40^{+0.07}_{-0.07}$ & (38, 39, 40) \\
    $6368\_25297$ & CAPERS-25297  & $9.9381 \pm 0.0003$  & $-16.37 \pm 1.65$ & $-1.43 \pm 0.06$ & $21.78^{+0.21}_{-0.26}$ & (41, 42, 43, 44) \\
    $2561\_37126$ & UNCOVER-37126& $9.85^{+0.03}_{-0.15}$ & $-20.07 \pm 0.08$ & $-2.74 \pm 0.06$ & $22.53^{+0.09}_{-0.09}$ & (45) \\
    $1210\_14177$ & JADES-z10-0 & $10.4^{+0.35}_{-0.10}$  & $-17.29 \pm 0.29$ & $-2.23 \pm 0.11$ & $19.67^{+1.20}_{-1.22}$ & (13, 14, 15) \\
    $2561\_13151^{(*)}$ & UNCOVER-13151 & $9.8026 \pm 0.0003$  & $-17.89 \pm 0.60$ & $-2.40 \pm 0.06$ & $21.35^{+0.22}_{-0.42}$  & (46, 47, 48, 44) \\
    $6368\_127376$ & CAPERS-127376 & $9.2848 \pm 0.0004$  & $-19.62 \pm 0.12$ & $-2.09 \pm 0.12$ & $22.12^{+0.23}_{-0.28}$ & \\
    $2756\_202^{(*)}$ & & $9.800 \pm 0.001$  & $-16.99 \pm 4.82$ & $-1.89 \pm 0.22$  & $19.42^{+1.08}_{-0.98}$  & \\
    $2561\_3686^{(*)}$ & UNCOVER-3686 & $9.3202 \pm 0.0004 $  & $-21.80 \pm 0.07$ & $-2.05 \pm 0.03$ & $21.58^{+0.11}_{-0.12}$ & (49, 50, 51, 41, 43, 46) \\
    $1181\_55757$ & JADES-GN-55757  & $9.7498 \pm 0.0006 $  & $-19.92 \pm 0.04$ & $-2.22 \pm 0.10$ & $18.90^{+0.74}_{-0.61}$  & (52, 43) \\
    $1286\_20088041$ & JADES-GS-20088041 & $9.712\pm 0.003$  & $-19.75 \pm 0.15$& $-2.50 \pm 0.21$  & $21.68^{+0.52}_{-0.72}$  & (53, 15)\\
    $1181\_59720$ & JADES-GN-59720 & $9.6356 \pm 0.0003$  & $-19.47 \pm 0.09$ & $-2.29 \pm 0.11$ & $19.28^{+0.96}_{-0.88}$ & (52) \\
    $4233\_833482$ & RUBIES-UDS-833482 & $9.3042 \pm 0.0003 $  & $-20.51 \pm 0.19$ & $-2.04 \pm 0.16$ & $21.29^{+0.62}_{-0.89}$ & (43) \\
    $2767\_11027^{(*)}$ & RXJ2129-11027 & $9.5127 \pm 0.0001 $ & $-16.66 \pm 3.65$ & $-1.26 \pm 0.18$ & $22.78^{+0.26}_{-0.30}$ & (54, 43)\\
    $1345\_80026$ & CEERS-80026 & $ 9.566 \pm 0.001$  & $-20.20 \pm 0.16$ & $-2.09 \pm 0.16$ & $22.25^{+0.29}_{-0.32}$ & (23)\\
    $2561\_22223^{(*)}$ & UNCOVER-22223 & $9.5704 \pm 0.0008 $ & $-17.17 \pm 0.33$ & $-2.37 \pm 0.12$ & $21.27^{+0.50}_{-1.00}$ & (43, 46)\\
    $6368\_87132$& CAPERS-87132 & $9.3833 \pm 0.0002 $  & $-18.46 \pm 0.12$& $-1.67 \pm 0.15$& $22.27^{+0.22}_{-0.27}$& (43, 52)\\
    $1181\_3990 $& JADES-GN-3990 & $9.3812 \pm 0.0002 $  & $-19.63 \pm 0.20$ & $-1.62 \pm 0.09$ & $19.72^{+1.09}_{-1.20}$ & (55, 43) \\
    $1181\_17858$ & JADES-GN-17858 & $9.2173 \pm 0.0003$  & $-19.86 \pm 0.08$ & $-2.17 \pm 0.14$ & $19.30^{+1.04}_{-0.85}$& (52) \\
    $1433\_3568^{(*)}$ & EBG-1 & $9.247 \pm 0.001$  & $-19.17 \pm 0.14$ & $-3.07 \pm 0.12$ & $19.53^{+1.04}_{-0.96}$ &  \\
    $6368\_22431 $& CAPERS-22431 & $9.2717 \pm 0.0001$  & $-20.51 \pm 0.02$ & $-2.54 \pm 0.04$ & $20.27^{+0.51}_{-1.39}$ & (43) \\
    $6368\_28597$ & CAPERS-28597 & $9.214 \pm 0.001$  & $-20.90 \pm 0.10$ & $-2.22 \pm 0.07$ & $21.27^{+0.30}_{-1.09}$ & \\
    $3215\_265801$& JADES-GS-265801 & $9.4437 \pm 0.0001$ & $-19.66 \pm 0.01$    & $-2.48 \pm 0.01$& $21.75^{+0.04}_{-0.04}$& (56, 57, 58, 59, 13, 41, 43) \\
    $4233\_918503$ & RUBIES-EGS-918503 & $9.2547 \pm 0.0004$ & $-20.51 \pm 0.36$ & $-2.11 \pm 0.20$ & $19.12^{+0.93}_{-0.80}$ & \\
    $1181\_19715$ & JADES-GN-19715 & $9.308 \pm 0.001$  & $-20.59 \pm 0.06$ & $-2.44 \pm 0.10$ & $21.68^{+0.32}_{-1.00}$ & (52) \\
    $4233\_952693$ & RUBIES-EGS-952693 & $9.0417 \pm 0.0003$  & $-20.40 \pm 0.13$ & $-1.97 \pm 0.10$ & $19.31^{+1.66}_{-0.92}$ & \\
    $1286\_20083087$ & JADES-GN-20083087 & $9.067 \pm 0.001$  & $-20.02 \pm 0.10$ & $-2.39 \pm 0.15$ & $19.82^{+1.23}_{-1.23}$ & \\
    $1181\_619$ & JADES-GN-619 & $9.0718 \pm 0.0003$ & $-20.29 \pm 0.10$ & $-2.21 \pm 0.08$ & $19.85^{+1.14}_{-1.26}$ & (52) \\
    $6368\_94554$ & CAPERS-94554 & $9.017 \pm 0.001$ & $-18.20 \pm 0.18$ & $-2.08 \pm 0.19$ & $19.35^{+1.06}_{-0.88}$ & \\
    $1286\_20077159$ & JADES-GS-20077159 & $9.0556 \pm 0.0002$  & $-20.27 \pm 0.11$ & $-1.71 \pm 0.12$ & $21.75^{+0.32}_{-0.52}$ & \\
\end{longtable}
\vspace{1cm}
\textbf{Notes: }$(^*)$ Lensed galaxies, UV magnitudes are corrected for lensing. As well as extensive overlap between the samples of \citet{Tang25} at $z>9$, and \citet{RobertsBorsani25} at $z>10$, individual references for discovery or analysis of specific galaxies follow: (1) \cite{Naidu25} (2) \cite{Schouws25} (3) \cite{Heintz25_z14} (4) \cite{Carniani25} (5) \cite{Carniani24a} (6) \cite{Scholtz25} (7) \cite{Ferrara24} (8) \cite{Helton25} (9) \cite{Wu25} (10) \cite{Qin25} (11) \cite{Cohon25} (12) \cite{Witstok25_LA} (13) \cite{Bunker23_jades} (14) \cite{CurtisLake23} (15) \cite{Hainline24a} (16) \cite{Hainline24b} (17) \cite{DEugenio24} (18) \cite{Castellano22} (19) \cite{Calabro24} (20) \cite{Ono23} (21) \cite{Zavala25} (22) \cite{Bakx23} (23) \cite{ArrabalHaro23} (24) \cite{Harikane23} (25) \cite{Harikane26} (26) \cite{Kokorev25} (27) \cite{Finkelstein23} (28) \cite{Finkelstein24} (29) \cite{Heintz24_DLA} (30) \cite{Witstok25_ALMA} (31) \cite{Oesch16} (32) \cite{Maiolino24} (33) \cite{Tacchella23} (34) \cite{AlvarezMarquez25} (35) \cite{Cameron23b} (36) \cite{Bunker23_gnz11} (37) \cite{Napolitano24} (38) \cite{Hsiao24a} (39) \cite{Hsiao24b} (40) \cite{Harikane24} (41) \cite{Nakane25} (42) \cite{McLeod24} (43) \cite{Pollock25} (44) \cite{Donnan25} (45) \cite{MarquesChaves26} (46) \cite{Fujimoto23_uncover} (47) \cite{Zitrin14} (48) \cite{RobertsBorsani23} (49) \cite{Boyett24b} (50) \cite{Atek23} (51) \cite{Yanagisawa25} (52) \cite{Tang24} (53) \cite{Donnan23} (54) \cite{Williams23} (55) \cite{Schaerer24} (56) \cite{Cameron23} (57) \cite{Curti25} (58) \cite{Boyett24} (59) \cite{Sanders25}.
\twocolumn

\begin{figure*}
    \centering
    \includegraphics[width=1.0\linewidth]{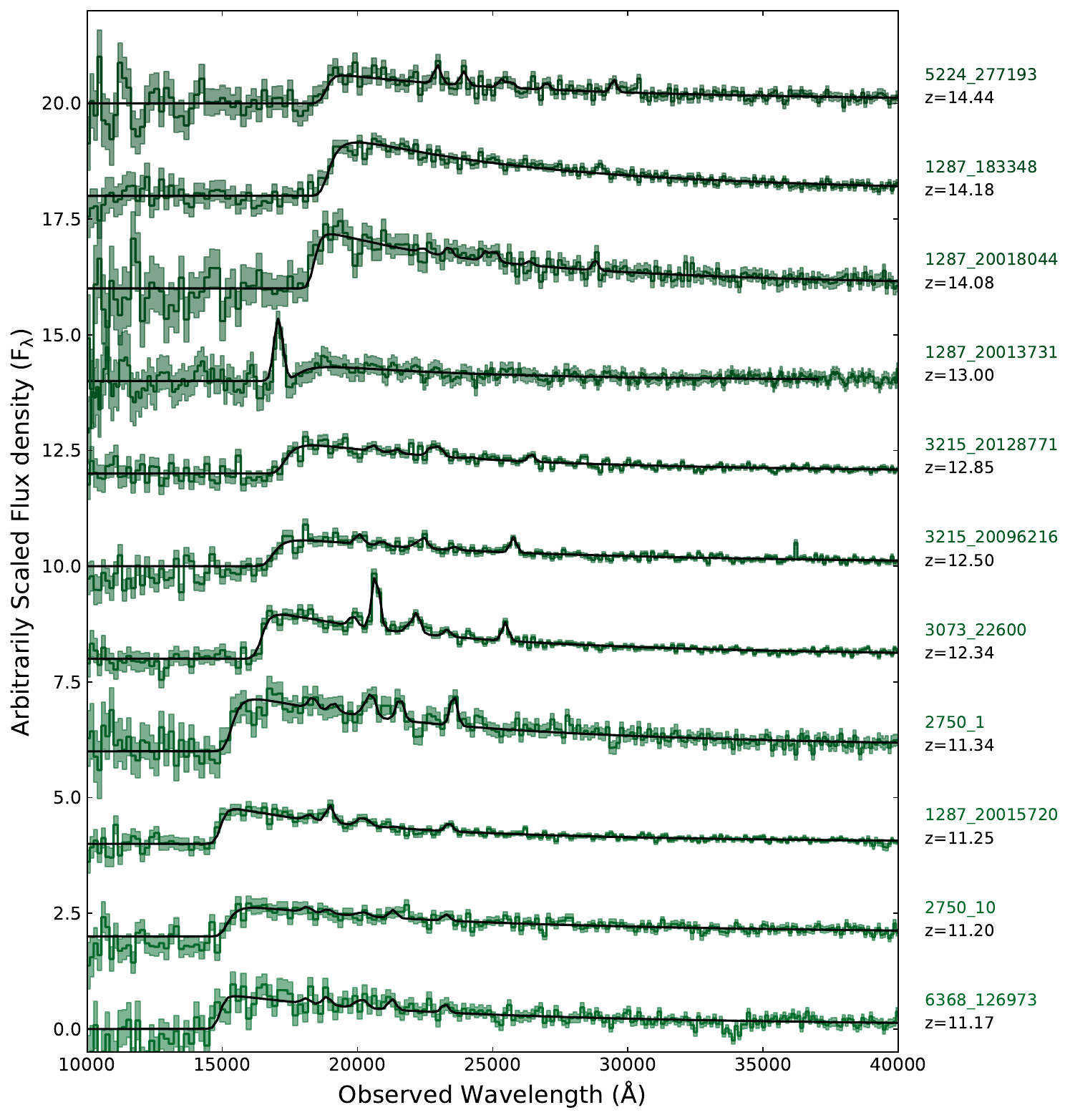}
    \caption{All $z>10$ spectra, showing UV turnover fitting. Note that when available, spectroscopic redshifts are measured from rest-optical emission lines.}
    \label{fig:z10A}
\end{figure*}

\begin{figure*}\ContinuedFloat
    \centering
    \includegraphics[width=1.0\linewidth]{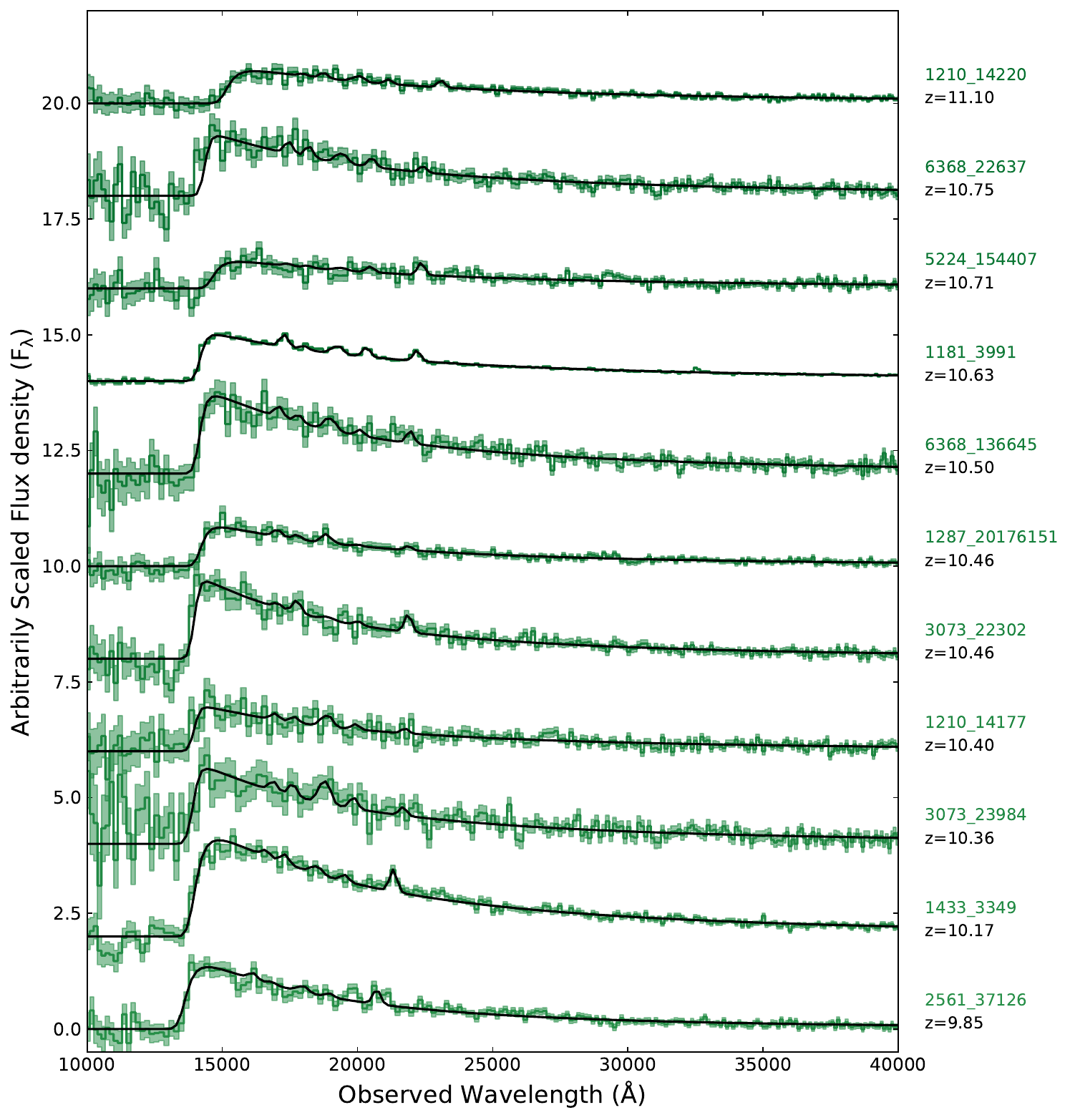}
    \caption{(continued)}
    \label{fig:z10B}
\end{figure*}

\begin{figure*}
    \centering
    \includegraphics[width=1.0\linewidth]{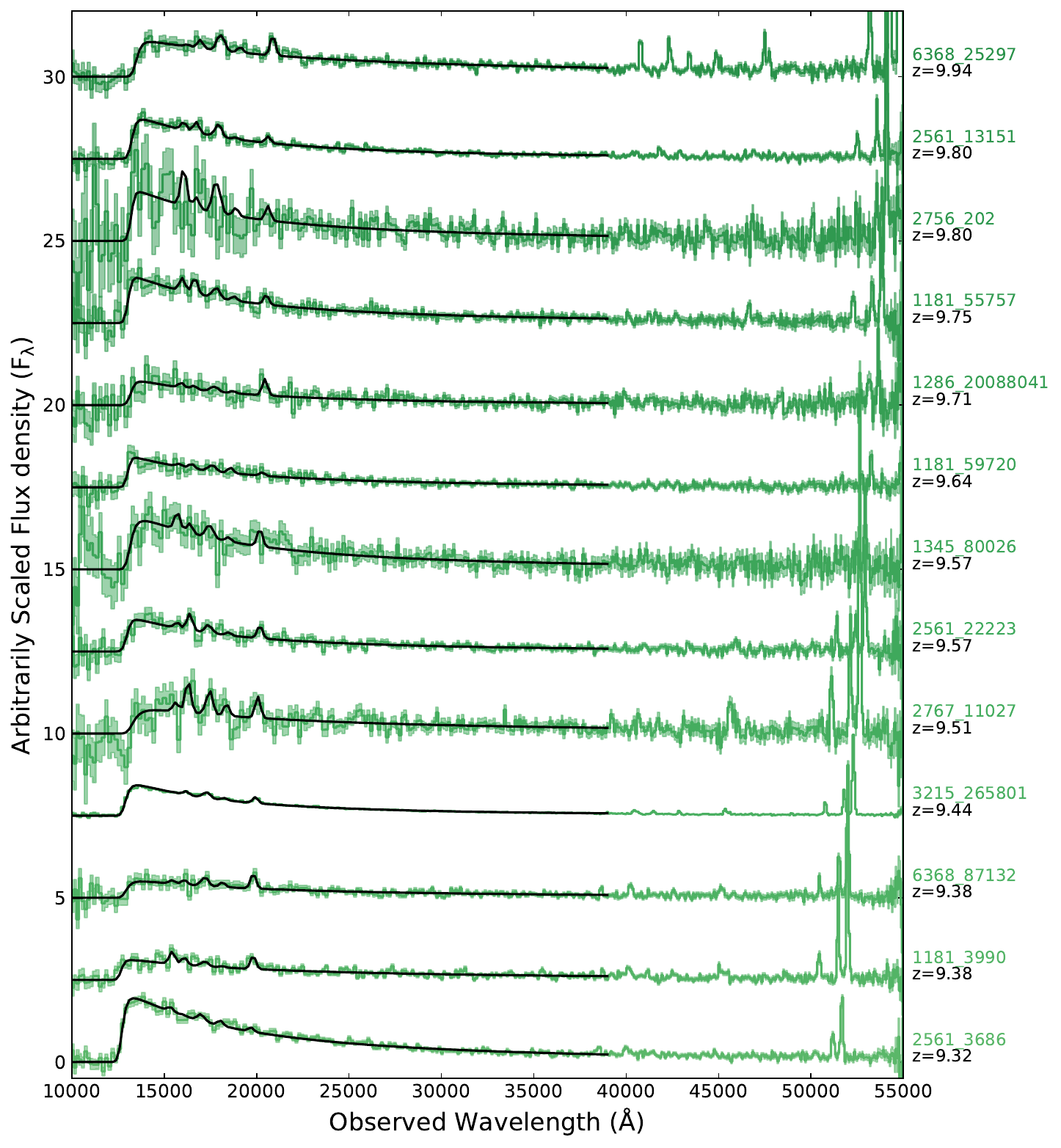}
    \caption{All $9<z<10$ spectra, showing UV turnover fitting. Note that in almost all cases, the best-fit spectroscopic redshift is derived from rest-optical emission lines.}
    \label{fig:z9A}
\end{figure*}

\begin{figure*}\ContinuedFloat
    \centering
    \includegraphics[width=1.0\linewidth]{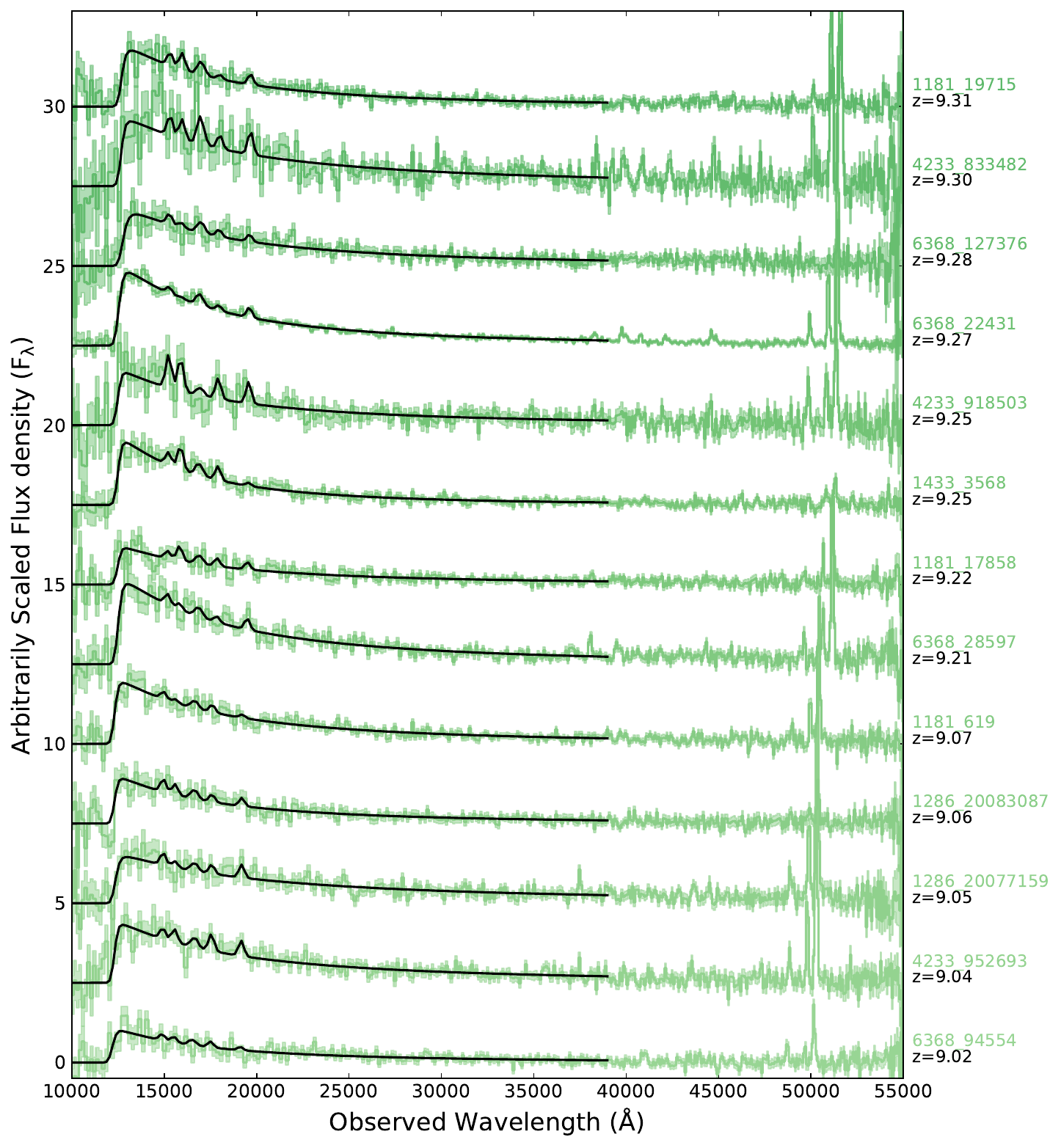}
    \caption{(continued)}
    \label{fig:z9B}
\end{figure*}
\end{document}